\documentclass[12pt,a4paper,epsf]{article}
\usepackage{graphics}
\usepackage{amssymb,amsmath}
\usepackage[dvips]{lscape,graphicx}
\usepackage{cite}

\newcommand{\ct}{\cite}
\newcommand{\lb}{\label}

\newcommand{\bc}{\begin{center}}
\newcommand{\ec}{\end{center}}
\newcommand{\bd}{\begin{displaymath}}
\newcommand{\ed}{\end{displaymath}}
\newcommand{\be}{\begin{equation}}
\newcommand{\ee}{\end{equation}}
\newcommand{\ba}{\begin{array}}
\newcommand{\ea}{\end{array}}
\newcommand{\bea}{\begin{eqnarray}}
\newcommand{\eea}{\end{eqnarray}}
\newcommand{\bt}{\begin{tabular}}
\newcommand{\et}{\end{tabular}}

\newcommand{\bp}{\begin{picture}}
\newcommand{\ep}{\end{picture}}
\newcommand{\bfi}{\begin{figure}}
\newcommand{\efi}{\end{figure}}

\begin{document}

\title{\Large\bf {Generalized Dual Symmetry 
 of Nonabelian Theories and the Freezing of ${\Large\bf \alpha_s}$}}

\author{
C.R.~Das ${}^{1}$ \footnote{\large\, crdas@imsc.res.in} ,
L.V.~Laperashvili ${}^{1,\, 2}$ \footnote{\large\, laper@itep.ru, laper@imsc.res.in} ,
H.B.~Nielsen ${}^{3}$ \footnote{\large\, hbech@nbi.dk} \\[5mm]
\itshape{${}^{1}$ The Institute of Mathematical Sciences, Chennai, India}\\[0mm]
\itshape{${}^{2}$ The Institute of Theoretical and Experimental Physics, Moscow, Russia}\\[0mm]
\itshape{${}^{3}$ The Niels Bohr Institute, Copenhagen, Denmark}}

\date{}
\maketitle

\begin{abstract}

The quantum Yang--Mills theory, describing a system of fields with non--dual (chromo--electric 
$g$) and dual (chromo--magnetic $\tilde g$) charges and revealing the generalized dual symmetry,
is developed by analogy with the Zwanziger formalism in QED. The renormalization group 
equations (RGEs) for pure nonabelian theories are analysed for both constants, 
$\alpha=g^2/4\pi$ and $\tilde\alpha={\tilde g}^2/4\pi$. The pure 
$SU(3)\times\widetilde {SU(3)}$ gauge theory is investigated as an example. We consider not 
only monopoles, but also dyons. The behaviour of the total $SU(3)$ $\beta$--function is
investigated in the whole region of $\alpha\equiv \alpha_s$: $0 \le \alpha < \infty$. It is 
shown that this $\beta$--function is antisymmetric under the interchange 
$\alpha \leftrightarrow 1/\alpha$ and is given by the well--known perturbative expansion 
not only for $\alpha \ll 1$, but also for $\alpha \gg 1$. Using an idea of the Maximal Abelian 
Projection by t' Hooft, we have considered the formation of strings -- the ANO flux tubes -- in 
the Higgs model of scalar monopole (or dyon) fields. In this model we have constructed the 
behaviour of the $\beta$--function in the vicinity of the point $\alpha = 1$, where it acquires 
a zero value. Considering the phase transition points at $\alpha\approx 0.4$ and 
$\alpha\approx 2.5$, we give the explanation of the freezing of $\alpha_s$. The evolution of 
$\alpha_s^{-1}(\mu)$ with energy scale $\mu$ and the behaviour of $V_{eff}(\mu)$ are 
investigated for both, perturbative and non--perturbative regions of QCD. It was shown 
that the effective potential has a minimum, ensured by the dual sector of QCD. 
The gluon condensate $F^2_0$, corresponding to this minimum, is predicted: 
$F^2_0\approx 0.15$ GeV$^4$, in agreement with the well--known results. 

\end{abstract}

\thispagestyle{empty}

\clearpage\newpage

\pagenumbering{arabic}

\section{Introduction}

In the last years gauge theories essentially operate with the fundamental idea of duality 
\ct{1,2,3,4,4i,4ii,4iii,5,5i,6,7,8,8i,9,9i,9ii,10,10i,11,12,13,14}.

Duality is a symmetry appearing in pure electrodynamics as invariance of the free Maxwell 
equations:
\be{\bigtriangledown}\cdot \vec {\bf B} = 0, \qquad
 {\bigtriangledown}\times \vec {\bf E} = - \partial_0\vec {\bf B},\lb{1}\ee 
\be {\bigtriangledown}\cdot \vec{\bf E} = 0, \qquad
 {\bigtriangledown}\times \vec{\bf B} = \partial_0\vec{\bf E},\lb{2}\ee 
under the interchange of electric and magnetic fields: 
\be\vec {\bf E} \to \vec {\bf B}, \qquad\vec{\bf B} \to  -\vec{\bf E}.\lb{3}\ee 

Letting 
\be F = \partial\wedge A = - (\partial\wedge B)^{*},\lb{4}\ee 
\be F^{*} = \partial\wedge B = (\partial\wedge A)^{*},\lb{5}\ee 
we see that the equations of motion: 
\be\partial_\lambda F_{\lambda\mu} = 0,\lb{6}\ee 
together with the Bianchi identity: 
\be\partial_\lambda F_{\lambda\mu}^{*} = 0,\lb{7}\ee 
are equivalent to Eqs.~(\ref{1}) and (\ref{2}), and show the invariance under the Hodge star 
operation on the field tensor: 
\be F_{\mu\nu}^{*} = \frac 12 \epsilon_{\mu\nu\rho\sigma} F_{\rho\sigma}.\lb{8}\ee 
This Hodge star duality, having a long history 
\ct{1,2,3,4,4i,4ii,4iii,5,5i,6,7,8,8i,9,9i,9ii,10,10i,11,12,13}, does not 
hold in general for nonabelian theories. In abelian theory Maxwell's equation (\ref{6}) is 
equivalent to the Bianchi identity for the dual field $F_{\mu\nu}^{*}$, which guarantees the
existence of a dual potential $B_{\mu}$ given by Eq.~(\ref{5}).

In the nonabelian theory, one usually starts with a gauge field $F_{\mu\nu}(x)$ derivable from a
potential $A_{\mu}(x)$: 
\be F_{\mu\nu} = \partial_{\nu}A_{\mu}(x) - \partial_{\mu}A_{\nu}(x) + i g [ A_{\mu}(x), 
A_{\nu}(x) ].\lb{9}\ee 
Considering only gauge groups with the Lie algebra of $SU(N)$, we have: 
\be A_{\mu}(x) = t^jA_{\mu}^j(x),\qquad  j = 1,...,N^2 - 1,\lb{10}\ee where $t^j$ are the 
generators of $SU(N)$ group. Equations of motion obtained by extremizing the corresponding 
action with respect to $A_{\mu}(x)$ gives: 
\be D_{\nu}F^{\mu\nu}(x) = 0,\lb{11}\ee 
where $D_{\mu}$ is the covariant derivative defined as 
\be D_{\mu} = \partial_{\mu} - ig[ A_{\mu}(x),...].\lb{12}\ee 

The analogy to electromagnetism is still rather close. But Yang--Mills equation (\ref{11}) does 
not imply in general the existence of a potential for the corresponding dual field
$F_{\mu\nu}^{*}$. This Yang--Mills equation itself can no longer be interpreted as the Bianchi 
identity for $F_{\mu\nu}^{*}(x)$, nor does it imply the existence of a dual potential ${\tilde
A}_{\mu}(x)$ satisfying 
\be F_{\mu\nu}^{*}(x) \stackrel{?}{=} \partial_{\nu}{\tilde A}_{\mu}(x) -
\partial_{\mu}{\tilde A}_{\nu}(x) + i{\tilde g} [ {\tilde A}_{\mu}(x),
{\tilde A}_{\nu}(x) ],\lb{13}\ee 
in parallel to (\ref{9}). This result means that the dual symmetry of the Yang--Mills theory 
under the Hodge star operation does not hold. So one has to seek a more general form of duality
for nonabelian theories than the Hodge star operation on the field tensor.

It was shown in \ct{9,9i,9ii,10,10i,11} that the classical Yang--Mills theory i
s symmetric under a 
generalized dual transform which reduces to the well--known electromagnetic duality of the 
abelian case.

\section{Loop space variables of nonabelian theories}

As in Refs.~\ct{9,9i,9ii,10,10i,11,12,13}, we investigate the $SU(N)$ nonabelian 
theories in terms of 
loop variables. For usual (non--dual) sector we consider the path ordered exponentials with 
closed loops (see Fig.~\ref{f1}): 
\be\Phi(C) = P\exp \left[ ig\oint_C A_{\mu}(\xi) d{\xi}^{\mu}\right] = P \exp \left[ 
ig\int_0^{2\pi}A_{\mu}(\xi)\dot {\xi}^{\mu}(s)ds\right], \lb{14}\ee  
where $C$ is a parameterized closed loop with coordinates ${\xi}^{\mu}(s)$ in the 
4--dimensional space. The loop is parameterized by $s$: $0\le s \le 2\pi$, and
\be\dot {\xi}^{\mu}(s)=\frac {d{\xi}^{\mu}(s)}{ds}. \lb{15}\ee
We also consider  the following unclosed loop variable \ct{15}:
\be\Phi(s_1, s_2)=P\exp\left[ig\int_{s_1}^{s_2} A_{\mu}(\xi) 
\dot {\xi}^{\mu}(s)ds\right].\lb{16}\ee
Therefore, $\Phi(C)\equiv \Phi(0, 2\pi)$.

For the dual sector we have (see Fig.~\ref{f2}):
\be\tilde\Phi(\tilde C) = P \exp \left[ i\tilde g\oint_{\tilde C} {\tilde A}_{\mu}(\eta)
d{\eta}^{\mu}\right] = P \exp \left[ i\tilde g\int_0^{2\pi} {\tilde A}_{\mu}(\eta)
\dot {\eta}^{\mu}(t) dt\right],\lb{17}\ee
where $\tilde C$ is a parameterized closed loop in the dual sector with coordinates 
$\eta^\mu(t)$ in the 4--dimensional space, and the loop parameter is $t$: $0\le t \le 2\pi$;
\be\dot\eta^\mu(t)=\frac{d\eta^\mu(t)}{dt}.\lb{18}\ee
The unclosed loop variables in the dual sector are:
\be\tilde \Phi(t_1, t_2) = P\exp \left[ i\tilde g\int_{t_1}^{t_2}\tilde A_\mu
(\eta) \dot\eta^\mu(t) dt\right].\lb{19}\ee
Therefore, $\tilde \Phi(\tilde C)\equiv \tilde \Phi(0, 2\pi)$. Here standard and dual sectors 
have coupling constants $g$ and $\tilde g$ respectively.

Considering (for simplicity of presentation) only gauge groups $SU(N)$, we have 
vector--potentials $A_{\mu}$ and $\tilde A_{\mu}$ belonging to the adjoint representation of 
$SU(N)$ and $\widetilde{SU(N)}$ groups:
\be A_{\mu}(x) = t^jA_{\mu}^j, \qquad \tilde A_{\mu}(x) = t^j\tilde
A_{\mu}^j, \qquad j=1,...,N^2 - 1. \lb{20} \ee

As a result, we consider nonabelian theories having a doubling of symmetry from $SU(N)$ to
\be SU(N) \times \widetilde {SU(N)}. \lb{21}\ee

\section{The nonabelian Zwanziger--type action and duality}

Following the idea of Zwanziger \ct{16,16i,17} (see also \ct{18,19}) to describe symmetrically 
non--dual and dual abelian fields $A_{\mu}$ and $\tilde A_{\mu}$, covariantly interacting with
electric $j_{\mu}^{(e)}$ and magnetic $j_{\mu}^{(m)}$ currents respectively, we suggest to 
construct the generalized Zwanziger formalism for the pure nonabelian gauge theories, 
considering the following Zwanziger--type action:
\bea S &=& - \frac 2K \int {\cal D}{\xi}^{\mu}ds \left\{ Tr(E^{\mu}[\xi|s]
     E_{\mu}[\xi|s]) + Tr ({\tilde E}^{\mu}[\xi|s] {\tilde E}_{\mu}[\xi|s])\right.\nonumber\\
&&+\left. i Tr(E^{\mu}[\xi|s]{\tilde E}_{\mu}^{(d)}[\xi|s]) + i Tr({\tilde E}^{\mu}[\xi|s] 
E_{\mu}^{(d)}[\xi|s])\right\} {\dot\xi}^{-2}(s) + S_{gf}.\lb{22}\eea 
Here we have used the Chan--Tsou variables \ct{9,9i,9ii,10,10i,11}:
\be E_{\mu}[\xi|s] = \Phi(s,0)F_{\mu}[\xi|s]{\Phi}^{-1}(s,0), \lb{23}\ee 
where 
\be F_{\mu}[\xi|s] = \frac {i}{g} {\Phi}^{-1}(C(\xi))\frac{\delta \Phi(C(\xi))}
{\delta \xi^{\mu}(s)}\lb{24}\ee 
are the Polyakov variables \ct{15}.

The illustration for the quantities $F_{\mu}[\xi|s]$ and $E_{\mu}[\xi|s]$ is given by 
Fig.~\ref{f3} and Fig.~\ref{f4}, considered in Refs.~\ct{9,9i,9ii,10,10i,11} 
(see also Appendix A). Using 
$\tilde \Phi$, we have the analogous expressions for ${\tilde F}_{\mu}[\xi|s]$ and 
${\tilde E}_{\mu}[\xi|s]$.

In Eq.~(\ref{22}) $K$ is the normalization constant: 
\be K = \int_0^{2\pi}ds{\Pi}_{s'\neq s}d^4\xi(s'),\lb{25}\ee 
and $S_{gf}$ is the gauge--fixing action: 
\be S_{gf} = \frac 2K \int {\cal D}\xi^{\mu} ds \left[ M_A^2 {(\dot{\xi}\cdot A)}^2 
+ M_B^2{({\dot\xi}\cdot \tilde A)}^2 \right]{\dot\xi}^{-2},\lb{26}\ee
which excludes ghosts in the theory \ct{18}. Also we have used a generalized dual operation 
\ct{9,9i,9ii,10,10i,11}:
\be E_{\mu}^{(d)}[\xi|s] = - \frac 2K \epsilon_{\mu\nu\rho\sigma}{\dot\xi}^{\nu}
  \int {\cal D}{\eta}^{\mu}dt \omega (\eta(t))E^{\rho}[\eta|t]
  {\omega}^{-1}(\eta(t)){\dot\eta}^{\sigma}(t){\dot\eta}^{-2}
  \delta (\eta(t) - \xi(s)).\lb{27}\ee 
The last integral in Eq.~(\ref{27}) is over all loops and over all points of each loop, and the 
factor $\omega(x)$ is just a rotational matrix allowing for the change of local frames between 
the two sets of variables. 

In the abelian case the expression (\ref{27}) coincides with the Hodge star operation:
\be F_{\mu\nu}^{*} = \frac 12 \epsilon_{\mu\nu\rho\sigma}F_{\rho\sigma}, \ee\lb{28}
but for nonabelian theories they are different.

From our Zwanziger--type action we have the following equations of motion: 
\be\frac{\delta E_{\mu}[\xi|s]}{\delta \xi^{\mu}(s)} = 0, \qquad
 \frac{\delta \tilde E_{\mu}[\xi|s]}{\delta \xi^{\mu}(s)} = 0. \lb{29}\ee 
Such a theory shows the invariance under the generalized dual operation (\ref{27}), e.g. has a 
dual symmetry under the interchange: 
\be E_{\mu} \longleftrightarrow  {\tilde E}_{\mu},\lb{30}\ee 
where ${\tilde E}_{\mu}$ is related with the generalized dual operation (\ref{27}): 
\be{\tilde E}_{\mu} = E^{(d)}_{\mu}.\lb{31}\ee

The regularization procedure considered in Refs.~\ct{9,9i,9ii,10,10i,11} 
leads to the following relations 
(see Appendix A):
\be\lim\limits_{\epsilon \to 0}E_{\mu}[\xi|s] = F_{\mu\nu}{\dot\xi}^{\nu}(s),\lb{32}\ee
\be\lim\limits_{\epsilon \to 0}{\tilde E}_{\mu}[\xi|s]
= {\tilde F}_{\mu\nu}{\dot\xi}^{\nu}(s).\lb{33}\ee
However,
\be\lim\limits_{\epsilon \to 0}E^{(d)}_{\mu}[\xi|s] \neq
- \frac 12 \epsilon_{\mu\nu\rho\sigma}{\dot\xi}_{\nu}F_{\rho\sigma},\lb{34}\ee
\be\lim\limits_{\epsilon \to 0}{\tilde E}^{(d)}_{\mu}[\xi|s] \neq
\frac 12 \epsilon_{\mu\nu\rho\sigma}{\dot\xi}_{\nu}{\tilde F}_{\rho\sigma},\lb{35}\ee
showing that for nonabelian theories the reduction to the Hodge star operation does not go 
through.

\section{The charge quantization condition}

Considering the Wilson operator:
\be A(C)=Tr\left(P\exp \left[ig\oint_C A_{\mu}(\xi) d{\xi}^{\mu}\right]\right), \lb{36}\ee 
which measures the chromo--magnetic flux through $C$ and creates the chromo--electric flux 
along $C$, and the dual operator: 
\be B(\tilde C)= Tr\left(P \exp \left[ i\tilde g\oint_{\tilde C} {\tilde
A}_{\mu}(\eta)d{\eta}^{\mu}\right]\right), \lb{37} \ee   
measuring the chromo--electric flux through $\tilde C$ and creating the chromo--magnetic flux
along $\tilde C$, we can use the t' Hooft commutation relation \ct{20}:
\be A(C)B(\tilde C) = B(\tilde C)A(C)\exp\left(\frac{2\pi in}{N}\right), \lb{38}\ee 
where $n$ is the number of times $\tilde C$ winds around $C$ and $N\ge 2$ is for the gauge 
group $SU(N)$. By this way, the authors of Refs.~\ct{9,9i,9ii,10,10i,11} 
have obtained the generalized 
charge quantization condition:
\be g\tilde g = 4\pi n,\qquad n\in Z, \lb{39}\ee 
which is called the Dirac--Schwinger--Zwanziger (DSZ) relation.

Using fine structure constants containing the elementary charges $g$ and $\tilde g$ (the case 
$n=1$ in Eq.~(\ref{39})):
\be\alpha = \frac{g^2}{4\pi},\qquad\tilde \alpha = \frac{{\tilde g}^2}{4\pi}, \lb{40}\ee 
we have the following charge quantization relation:
\be \alpha \tilde \alpha = 1.\lb{41} \ee

\section{Renormalization group equations and duality}

For pure nonabelian gauge theories, generalized duality gives a symmetry under the interchange: 
\be \alpha \leftrightarrow \tilde \alpha, \lb{42} \ee
or according to the relation (\ref{41}): 
\be \alpha \leftrightarrow \frac {1}{\alpha}. \lb{43} \ee 
For the first time such a symmetry was considered by Montonen and Olive in Ref.~\ct{21}.

In nonabelian  theories with chromo--electric and chromo--magnetic charges, the derivatives 
$d\ln\alpha/dt$ and $d\ln\tilde \alpha/dt$ are only a function of the effective constants 
$\alpha$ and $\tilde \alpha$, as in the Gell--Mann--Low theory \ct{22}. Here 
\be t=\ln\left(\frac{\mu^2}{M^2}\right),\lb{44}\ee 
$\mu $ is the energy variable and $M$ is the renormalization scale.

In Refs.~\ct{19,23,29} and in Appendix B
it was shown  that when we consider the both, chromo--electric and 
chromo--magnetic (non--dual and dual), charges we can write the following general expressions 
for $\beta$--functions of the renormalization group equations (RGEs):
\be \frac{\mbox{d}\ln \alpha(t)}{\mbox{dt}} = -\frac{\mbox{d}\ln {\tilde \alpha}(t)}{\mbox{dt}}
= \beta(\alpha ) - \beta(\tilde \alpha) = \beta^{(total)}(\alpha),\lb{45}\ee 
where our $\beta(\alpha )$ for $\alpha\ll 1$ coincides with the perturbative $\beta$--function, 
which is well--known in literature, for example, for $SU(3)$ gauge theory. Eq.~(\ref{45}) is a 
consequence of the dual symmetry and the charge quantization condition valid for arbitrary $t$:
\be\alpha(t) \tilde \alpha (t) = 1.\lb{46}\ee
Here we see that the total $\beta$--function:
\be \beta^{(total)}(\alpha) = \beta(\alpha) - \beta(\tilde\alpha)\lb{47}\ee
is {\underline {antisymmetric}} under the interchange
\be\alpha \leftrightarrow \tilde \alpha,\qquad {\mbox{or}} \qquad\alpha \leftrightarrow 
\frac 1{\alpha},\lb{48}\ee
what means that $\beta^{(total)}(\alpha)$ has a zero at the point $\alpha = \tilde \alpha = 1$:
\be\beta^{(total)}(\alpha=\tilde \alpha = 1) = 0. \lb{49}\ee

\section{An example of $\beta$--function for the pure $SU(3)$ colour gauge group 
(Part I)}

The investigation of gluondynamics -- the pure $SU(3)$ colour gauge theory -- shows that at 
sufficiently small $\alpha < 1$ the $\beta$--function in the 3--loop approximation is given by 
the following series over $\alpha /4\pi$ \ct{24}:
\be \beta(\alpha ) = - \left[\beta_0\frac{\alpha}{4\pi} +\beta_1{\left(\frac{\alpha}
{4\pi}\right)}^2 + \beta_2{\left(\frac{\alpha}{4\pi}\right)}^3 + ...\right],\lb{50}\ee 
where for gluondynamics we have:
\be\beta_0 = 11,\qquad \beta_1 = 102, \qquad \beta_2 = 1428.5,\lb{51}\ee 
and for QCD:
\be\beta_0 = 11 - {\frac 23}N_f,\qquad \beta_1 = 102 - \frac{38}{3}N_f, \qquad 
\beta_2 = 1428.5 - \frac{5033}{18}N_f + \frac{325}{54}N_f^2.\lb{52}\ee 

It is very important that QCD shows a phenomenon of the freezing of $\alpha \equiv \alpha_s$ 
at the point $\alpha\approx 0.4$ (see Refs.~\ct{25,25i,25ii,26,26i}). 
This idea has an explanation by 
string formation: for $\alpha_s > 0.4$ we have the confinement of chromo--electric charges by 
chromo--electric flux tubes -- ANO strings \ct{27,28}. Then the chromo--electric charge becomes 
almost unchanged, what means that in the region of confinement $\beta(\alpha_s)\approx 0$. Such 
a phenomenon also was considered in Ref.~\ct{29} and in the review \ct{30}. 

For  $\tilde \alpha_s > 0.4$ we have the confinement of chromo--magnetic charges by 
chromo--magnetic ANO flux tubes. 

Assuming the freezing of the QCD coupling constants, we have:
\be \beta(\alpha) = 0 \qquad {\mbox{for}}\qquad \alpha > 0.4,\lb{53}\ee
and (by dual symmetry):
\be \beta(\tilde \alpha) = 0 \qquad {\mbox{for}}\qquad \tilde \alpha > 0.4.\lb{54}\ee 
Taking into account the condition $\alpha \tilde \alpha = 1$, we see that the value 
$\tilde \alpha = 0.4$ corresponds to the point $\alpha = 2.5$. As a result, the region of the 
confinement of chromo--electric and chromo--magnetic charges is given by the following 
requirement:
\be\beta(\alpha)^{(total)} = 0 \qquad {\mbox{for}}\qquad 0.4 < \alpha< 2.5.\lb{55}\ee
The behaviour of  $\beta^{(total)}(\alpha)$, given by Eqs.~(\ref{45}), (\ref{50}), (\ref{51}) 
and (\ref{55}), is shown in Fig.~\ref{f1} for the case of the pure 
$SU(3)\times\widetilde{SU(3)}$ gauge theory.

\section{The ``abelization" of monopole vacuum in nonabelian 
gauge theories}

\subsection{Maximal Abelian Projection  method}

In the light of contemporary ideas of the abelization of $SU(N)$ gauge theories \ct{31,32} 
(see also the review \ct{33} and Refs.~\ct{34,35,36,37,38}), it seems attractive to carry out 
the following speculations concerning to the behaviour of $\beta^{(total)}(\alpha)$ in the 
vicinity of the point $\alpha = 1$.

As it follows from the lattice investigations of pure $SU(3)$ theories \ct{32}, in some region 
of $\alpha > \alpha_{conf}$, gauge field $A_{\mu}^J$ $(J=1,...8)$ makes up composite 
configurations of monopoles which form a monopole condensate creating strings between the 
chromo--electric charges, according to scenarios given in Refs.~\ct{1,2}.

It is natural to think that the same configurations are created in the local $SU(3)$ gauge 
theory and imagine them as {\it the Higgs fields $\tilde \phi (x)$ of scalar chromo--magnetic 
monopoles.} Such investigations 
(see Refs.~\ct{39,39i,40,41,42,43,43i,44,44i,45,46,47,48,49,50}) were 
performed and their phenomenological predictions are quite successful.

In Ref.~\ct{31} t' Hooft developed a method of the Maximal Abelian Projection (MAP) suggested 
to consider such a gauge, in which monopole degrees of freedom, hidden in composite monopole 
configurations, become explicit and abelian. According to this method, scalar monopoles 
interact only with diagonal $SU(3)$ components of gauge fields ${({\tilde A}_{\mu})}^i_j$ (here
$i,j = 1,2,3$ are color indices). Non--diagonal $SU(3)$ components of gauge fields are 
suppressed and, as it was shown in Ref.~\ct{31} and \ct{33,34,35,36,37,38}, the interaction of 
monopoles with dual gluons is described by $U(1)\otimes U(1)$ subgroup of $SU(3)$ group 
\ct{31,34}. In general, we have $U(1)^{N-1}\subset SU(N)$ for $SU(N)$ gauge theory and $N-1$ 
types of monopoles. In Appendix C we present the formal procedure for the Maximal Abelian 
Projection method in continuum $SU(N)$ gluondynamics, following the review \ct{33}. 

The vacuum abelization of $SU(N)$ gauge theories is quite attractive to consider the behaviour 
of the $SU(3)$ total $\beta$--function in the vicinity of the point $\alpha = 1$.

According to the MAP, scalar monopoles are created in the non--perturbative region only by 
diagonal $SU(3)$ components ${({A}_{\mu})}^i_i$ of gauge fields ${({A}_{\mu})}^i_j$, and 
interact only with diagonal $\widetilde {SU(3)}$ components of gauge fields 
${({\tilde A}_\mu)}^i_j$.

In the non--perturbative region, non--diagonal $SU(3)$ and $\widetilde {SU(3)}$ components of 
gauge fields are suppressed and the interaction of monopoles with dual gluons is described by
$\widetilde {U(1)}\otimes \widetilde {U(1)}$ (Cartan) subgroup of $\widetilde {SU(3)}$ group.
These monopoles can be approximately described by the Higgs fields $\tilde \phi(x)$ of scalar 
chromo--magnetic monopoles interacting with gauge fields $\tilde A_{\mu}^J$.

Recalling the generalized dual symmetry, we are forced to assume that similar composite 
configurations have to be produced by dual gauge fields $\tilde A_{\mu}^J$, and described by 
the Higgs fields $\phi(x)$ of scalar chromo--electric ``monopoles" interacting with gauge 
fields $A_{\mu}^J$. The interaction of ``monopoles'' with gluons also is described by 
$U(1)\otimes U(1)$ subgroup of $SU(3)$ group. In general, we have $N-1$ types of ``monopoles''
belonging to the subgroup $U(1)^{N-1}\subset SU(N)$.

\subsection{$SU(3)$ gauge theory: field equations for monopoles and dyons}

The generators of the Cartan subgroup are given by the following diagonal Gell--Mann matrices:
\be t^3 = \frac{\lambda^3}{2} \qquad {\mbox{and}}\qquad t^8 = \frac{\lambda^8}{2},\lb{56}\ee
and in the non--perturbative region of $SU(3)$ gauge theory we have the following equations for 
diagonal $F_{\mu\nu}$, $\phi$ and $\tilde \phi$: 
\be\partial_{\nu}F_{\mu\nu}^{J=3,8} =\frac{i}{2} g\left[\phi^{+}\left(\frac{\lambda^{3,8}}
{2}\right){\cal D}_{\mu}\phi - ({\cal D}_{\mu}\phi)^+\left(\frac{\lambda^{3,8}}{2}\right)
\phi\right],\lb{57}\ee 
and 
\be\partial_{\nu}{\tilde F}_{\mu\nu}^{J=3,8} = \frac{i}{2}\tilde g\left[\tilde \phi^{+}\left
(\frac{\lambda^{3,8}}{2}\right){\tilde {\cal D}}_{\mu}\tilde\phi - ({\tilde {\cal D}}_{\mu}
\tilde\phi)^{+}\left(\frac{\lambda^{3,8}}{2}\right)\tilde\phi\right],\lb{58}\ee 
where 
\be {\cal D}_{\mu} = \partial_{\mu} - ig[A_{\mu},...]\qquad {\mbox{and}}\qquad
{\tilde {\cal D}}_{\mu} = \partial_{\mu} - i{\tilde g}[{\tilde A}_{\mu},...].\lb{59}\ee
We can choose two independent abelian monopoles as:
\be {\tilde \phi}_1 = ({\tilde \phi})^1_1 \qquad {\mbox{and}}
\qquad {\tilde \phi}_2 = ({\tilde \phi})^2_2,\lb{60}\ee 
and two independent abelian scalar fields with electric charges as:
\be{\phi}_1 = ({\phi})^1_1 \qquad {\mbox{and}}\qquad {\phi}_2 = ({\phi})^2_2.\lb{61}\ee

Considering the radiative corrections to the gluon propagator (see Fig.~\ref{f6}), we see that 
both abelian monopoles ${\tilde \phi}_1$ and ${\tilde \phi}_2$ have the monopole charge 
${\tilde g}_{(MAP)}$, however, both abelian ``monopoles'' $\phi_1$ and $\phi_2$ acquire the 
electric charge $g_{(MAP)}$. It was shown in Ref.~\ct{46} (see also the review \ct{30}) that, as
a result of the averaging over MAPs, near the critical point we have the following approximate 
relation between the charge of the abelian scalar particle, belonging to the Cartan 
$U(1)\times U(1)$ algebra, and $SU(N)$ coupling constant:
\be\alpha_N\approx \frac{N}{2}\sqrt{\frac{N+1}{N-1}}\alpha_{U(1)}.\lb{62}\ee
In the case of $SU(3)$ gauge theory, we have:
\be\alpha_3\approx \frac {3}{\sqrt 2}\alpha_{U(1)},\lb{63}\ee
what gives the following result:
\be\alpha_{(MAP)} \approx 0.5\alpha\qquad  
{\mbox{and}}\qquad  {\tilde \alpha}_{(MAP)} \approx 0.5\tilde \alpha.\lb{64}\ee
Using notations: 
\be f_{\mu\nu,i}\equiv (F_{\mu\nu})^i_i,\qquad a_{\mu,i}\equiv (A_{\mu})^i_i
\qquad {\mbox{and}}\qquad {\tilde a}_{\mu,i}\equiv ({\tilde A}_{\mu})^i_i,\lb{65}\ee
we have the following equations valid into the non--perturbative region of QCD 
$(i=1,2)$: 
\be\partial_{\nu}f_{\mu\nu,i} =\frac{i}{2} g_{(MAP)}[\phi_i^{+}{\cal D}_{\mu}\phi_i -
({\cal D}_{\mu}\phi_i)^{+}\phi_i],\lb{66}\ee
and
\be\partial_{\nu} f^{*}_{\mu\nu,i} = \frac{i}{2}{\tilde g}_{(MAP)}[{\tilde\phi}_i^{+}
{\tilde {\cal D}}_{\mu}{\tilde \phi}_i - ({\tilde {\cal D}}_{\mu}{\tilde\phi}_i)^{+}
{\tilde \phi}_i]. \lb{67}\ee 

A dual symmetry of pure nonabelian theories leads to the natural assumption that in the 
non--perturbative region, not monopoles and ``monopoles", but dyons are responsible for the
confinement. It was shown in Refs.~\ct{51,51a,51ai,51b,52} 
that color confinement in QCD is caused
by dyon condensation: the QCD vacuum is a media of condensed dyons. Then the MAP method leads
to the abelian Higgs model of dyons, which are described by united abelian scalar fields 
$\phi_{1,2,...N-1}$ having simultaneously electric and magnetic charges, and we have the 
following field equations for each components $i=1,2,..., N-1$:
\be\partial_{\lambda} f_{\lambda\mu,i} = ig_{(MAP)}[\phi^+_i{\cal D}_{\mu}\phi_i - 
({\cal D}_{\mu}\phi_i)^+ \phi_i], \lb{68} \ee
and 
\be\partial_{\lambda} f_{\lambda\mu,i}^{*} = i{\tilde g}_{(MAP)}[\phi^+_i{\tilde 
{\cal D}}_{\mu} \phi_i - ({\tilde {\cal D}}_{\mu}\phi_i)^+ \phi_i ],\lb{69}\ee 
where 
\be{\alpha}_{(MAP)}\,\,({\mbox{or}}\,\,{\tilde \alpha}_{(MAP)})\approx 
\frac{2}{N}\sqrt{\frac{N-1}{N+1}}\alpha_N\,\,({\mbox{or}}\,\,\tilde \alpha_N).\lb{70}\ee 
But the theory of dyons needs an additional investigation.

\section{$\beta$--function in the case of the pure $SU(3)$ colour gauge group 
(Part II)}

\subsection{$\beta$--function of scalar electrodynamics}

In the case of scalar electrodynamics, which is an abelian ($A$) gauge theory, we have the
following $\beta$--function in the two--loop approximation \ct{53,53i,53ii,54,54i}: 
\be \beta_A(\alpha^{(em)}) = \frac{\alpha^{(em)}}{12\pi}\left(1 + 3\frac{\alpha^{(em)}}
{4\pi} + ...\right).\lb{71}\ee 
For this abelian theory we have the Dirac relation:
\be \alpha^{(em)} {\tilde \alpha}^{(em)} = \frac 14, \lb{72}\ee
and the following RGEs for electric and magnetic fine structure constants:
$$\frac{\mbox{d}\ln \alpha^{(em)}(t)}{\mbox{d}t} = - \frac{\mbox{d}\ln {\tilde 
\alpha}^{(em)}(t)}{\mbox{d}t} = \beta_A(\alpha^{(em)}) - \beta_A({\tilde \alpha}^{(em)})$$
\be = \frac{\alpha^{(em)} - {\tilde \alpha}^{(em)}}{12\pi}\left(1 + 3\frac{\alpha^{(em)} + 
{\tilde \alpha}^{(em)}}{4\pi} + ...\right).\lb{73}\ee

As it was shown in Ref.~\ct{19}, the last RGEs can be considered simultaneously by perturbation 
theory only in the small region:
\be 0.2\lesssim \alpha^{(em)}, {\tilde \alpha}^{(em)} \lesssim 1. \lb{74}\ee
These approximate inequalities are valid for all abelian theories.

\subsection{Phase transition couplings for scalar electrodynamics}

The behaviour of the effective fine structure constants was investigated in the vicinity of the 
phase transition point in compact lattice QED by the Monte Carlo 
simulation method \ct{55,55i,56}.
The following result was obtained:
\be \alpha_{crit}^{lat.QED}\approx 0.20\pm 0.015, \qquad
\tilde \alpha_{crit}^{lat.QED}\approx 1.25\pm 0.10,\lb{75}\ee 
which is very close to the perturbative region (\ref{74}) for constants $\alpha^{(em)} $ and 
${\tilde \alpha}^{(em)} $. Using the two--loop approximation for the effective potential in the 
Higgs model of dual scalar electrodynamics, we have obtained 
in Refs.~\ct{44,44i,45,46,47,48,49} 
and \ct{56a} the following result:
\be \alpha_{crit}^{(em)}\approx 0.21,\qquad{\tilde \alpha}_{crit}^{(em)}\approx 1.20.\lb{76}\ee 
These values also are very close to the above--mentioned region (\ref{74}) of the abelian 
theory when both, dual and non--dual, charges are perturbative.

\subsection{Freezing of $\alpha_s$}

According to results of the previous Subsection, our abelian monopoles (or dyons), arising in 
QCD as a result of MAP, have the following critical dual constant value: 
\be {\tilde \alpha}_{(MAP)}^{(crit)}\approx 1.25,\lb{77}\ee 
what gives the beginning of the confinement region in $SU(3)$ gluondynamics (and QCD):
\be \alpha_1 = \alpha_{conf}= \frac 1{{\tilde\alpha}^{(crit)}} \approx
\frac 1{2{\tilde \alpha}_{(MAP)}^{(crit)}}\approx \frac 1{2.5} = 0.4.\lb{78}\ee 

We have received an explanation of the freezing value of $\alpha\equiv \alpha_s$.
By dual symmetry, the end of the perturbative region for scalar field $\phi$ is:
\be{\tilde \alpha}_{conf} \approx 0.4, \lb{79}\ee 
what corresponds to
\be\alpha_2 = \frac 1{\alpha_1}\approx 2.5. \lb{80}\ee

\subsection{$\beta$--function for the pure $SU(3)$ gauge theory}

The investigation, given by the previous Subsection, shows that in the region: 
\be 0.4\lesssim \alpha, \tilde{\alpha} \lesssim 2.5\lb{81}\ee 
we have an abelian theory (``abelian dominance'') with two scalar monopole fields 
${\tilde \phi}_{1,2}$ and two scalar electric fields ${\phi}_{1,2}$. The corresponding 
$\beta$--functions are: 
$$\frac{\mbox{d}\ln {\alpha}_{(MAP)}(t)}{\mbox{d}t} = - \frac{\mbox{d}\ln {\tilde 
\alpha}_{(MAP)}(t)}{\mbox{d}t} = \beta_A(\alpha_{(MAP)}) - \beta_A({\tilde \alpha}_{(MAP)})$$
\be = 2\left[\frac{{\alpha}_{(MAP)} - {\tilde \alpha}_{(MAP)}}{12\pi}\left(1 + 
3\frac{{\alpha}_{(MAP)} + {\tilde\alpha}_{(MAP)}}{4\pi} + ...\right)\right],\lb{82}\ee 
what gives the following $\beta$--functions, according to Eq.~(\ref{64}): 
$$ \frac{\mbox{d}\ln \alpha(t)}{\mbox{d}t} = - \frac{\mbox{d}\ln \tilde \alpha (t)}{\mbox{d}t}
= \beta_A(\alpha) - \beta_A(\tilde \alpha)$$
\be \approx \frac{\alpha - \tilde \alpha}{12\pi} \left(1 + 3\frac{\alpha + 
\tilde \alpha}{8\pi} + ...\right),\lb{83}\ee 
valid in the region (\ref{81}). In Eq.~(\ref{83}) we have:
\be\beta_A(\alpha)\approx\frac{\alpha}{12\pi}\left(1+3\frac{\alpha}{8\pi}+...\right),\lb{84}\ee 
and
\be \beta_A(\tilde \alpha)\approx \frac{\tilde \alpha}{12\pi}
\left(1 + 3\frac{\tilde \alpha}{8\pi} + ...\right).\lb{85}\ee 

The behaviour of the total $\beta$--function for the pure $SU(3)\times \widetilde{SU(3)}$ colour
gauge theory is given by Fig.~\ref{f7}, where the curve 1 describes the contribution of usual 
gluons for $\alpha < 0.4$ (see (a) of Fig.~\ref{f8}), but a tail of $\beta^{total}(\alpha )$, 
corresponding to $\alpha > 2.5$, is described by the curve 2, which presents loop contributions 
of diagrams in (b) of Fig.~\ref{f8}. The curve 1' describes the perturbative QCD 
$\beta$--function with quark and gluon contributions. The curve 3 presents a sum of 
contributions of scalar ``monopoles'' given by function $\beta_A(\alpha)$, and scalar monopoles 
described by function $-\beta_A(1/\alpha)$. Both of them exist in the non--perturbative region 
of gluondynamics, or QCD. The critical points: $\alpha_1\approx 0.4$ and $\alpha_2\approx 2.5$
also are shown in Fig.~\ref{f7}. Of course, we do not know the behaviour of the total 
$\beta$--function near the phase transition points. But these points explain an approximate 
freezing of $\alpha$ in the region (\ref{81}), where both charges, chromo--electric and 
chromo--magnetic ones, are confined. Chromo--electric strings exist for $\alpha > 0.4$, and 
chromo--magnetic ones exist for $\alpha < 2.5$. The region of strings is shown in Fig.~\ref{f7}.
Also we see that the total $\beta$--function has a zero at the point $\alpha=\tilde \alpha=1$,
predicted by our $SU(3)\times \widetilde{SU(3)}$ gauge theory. The behaviour of the total 
$\beta$--function, given by Fig.~\ref{f7}, is valid in the case of dyons, for which we have the 
same RGEs.

The ideas of Refs.~\ct{11,12} are also valid in the case of the Family replicated gauge group 
models (see, for example, Refs.~\ct{57,58} and the review \ct{59,59i}), 
where magnetic charges of 
monopoles (or dyons) are essentially diminished in comparison with those of the SM. We have left
these models for future investigations. 

In the pure $SU(3)\times \widetilde{SU(3)}$ gauge theory
there is no region of $\alpha $ when the perturbative expansions over $g$ and $\tilde g$ exist
simultaneously: when the non--dual sector is unconfined, then the dual sector is entirely 
confined and vice versa. Such a situation takes place also in the case of non--abelian theories
with matter fields.

\section{Nonabelian  theories  with matter fields}

Let us consider now nonabelian theories with matter fields having charge $ng$, or dual charge 
$n\tilde g$ (monopoles), or both of them (dyons).

If a surface $\Sigma $ in spacetime is parameterized as a closed loop in loop space, then via 
Eq.~(\ref{14}) it corresponds to a closed loop $\Gamma_{\Sigma}$ in the gauge group $G_g$. We 
say that the surface $\Sigma $ encloses a monopole if $\Gamma_{\Sigma}$ is in a non--trivial
homotopy class of $G_g$. This generalizes the Dirac magnetic monopole \ct{60} to the 
nonabelian case.

If matter fields, both non--dual and dual, exist in the nonabelian gauge theory, then a total 
system of fields is described by the action having the following structure:
\be S^{(total)} = S +  S_{(m)}+  S_{(d.m.)},\lb{86}\ee 
where $S$ is the Zwanziger--type action (\ref{22}) for gauge fields, $S_{(m)}$ is an action of 
matter fields, and $S_{(d.m.)}$ describes dual matter fields (monopoles). For dyons we have:
\be S^{(total)} = S +  S_D,\lb{87}\ee 
where $S_D$ is an action of dyon matter fields. Then (apart of dyons): 
\be S^{(total)} \approx S^{(n/d)} + S_{(m)}\qquad \mbox{for}\qquad \alpha < 1, \lb{88}\ee 
and
\be {\tilde S}^{(total)} \approx S^{(d)} +  S_{(d.m.)}\qquad \mbox{for}\qquad \alpha > 
1,\lb{89}\ee
where $S_{(n/d,d)}$ are non--dual and dual actions of gauge fields.

Nonabelian theories, revealing the (generalized) dual symmetry, have the following properties:

\begin{enumerate}
\item[1.] Monopoles of $A_{\mu}$ are charges of ${\tilde A}_{\mu}$, and ``monopoles'' of 
$\tilde A_{\mu}$ are charges of $A_{\mu}$.
\item[2.] If monopoles, as well as the charged particles, are Dirac fermions, then they are 
described by the Dirac Lagrangians:
\be L_{(m)} = {\bar \psi}\gamma_{\mu}(iD_{\mu} - m)\psi,\lb{90}\ee
\be L_{(d.m.)} = {\bar \chi}\gamma_{\mu}(i{\tilde D}_{\mu}- \tilde m)\chi,\lb{91}\ee
where covariant derivatives $D_{\mu}$ and ${\tilde D}_{\mu}$ are given by Eqs.~(\ref{59}).

The action of matter fields is:
\be S_{(m)\qquad\mbox{or}\qquad(d.m.)} = \int d^4x L_{(m)\qquad\mbox{or}\qquad(d.m.)}.\lb{92}\ee
\item[3.] Charged particles and monopoles can be the Klein--Gordon complex scalars:
\be L_{(m)} = \frac 12 [{|D_{\mu}\phi |}^2 - m^2 {|\phi|}^2],\lb{93}\ee
\be L_{(d.m.)} = \frac 12 [{|{\tilde D}_{\mu}\tilde \phi |}^2 -{\tilde m}^2 
{|\tilde \phi|}^2], \lb{94}\ee
or Higgs scalars:
\be L_{(m)} = \frac 12 [{|D_{\mu}\phi |}^2 - U(|\phi |)],\lb{95}\ee
\be L_{(d.m.)} = \frac 12 [{|{\tilde D}_{\mu}\tilde \phi |}^2 - U(|\tilde \phi |)],\lb{96}\ee
where
\be U(|\phi |) = \frac 12 m^2 {|\phi |}^2 +\frac {\lambda}4 {|\phi |}^4,\lb{97}\ee
and
\be U(|\tilde \phi |) = \frac 12 {\tilde m}^2 {|\tilde \phi |}^2 +
\frac {\tilde \lambda}4 {|\tilde \phi |}^4\lb{98}\ee
are the Higgs potentials.

All these matter fields can belong to different (for example, fundamental or adjoint) 
representations of $SU(N)$ group. In Refs.~\ct{9,9i,9ii,10,10i,11} 
monopoles belong to the fundamental 
representation of $SU(N)$.

For dyons -- particles having both, electric and magnetic, charges simultaneously -- we have in 
Eqs.~(\ref{90}-\ref{98}):
\be\psi^{ap}\equiv \chi^{ap}\lb{99}\ee 
and
\be\phi^{ap}\equiv \tilde \phi^{ap},\lb{100}\ee 
where $a,p$ are non--dual and dual indices of the $SU(N)$ and $\widetilde {SU(N)}$ 
representations respectively.
\item[4.] The charges of matter fields $g_m$ and ${\tilde g}_m$ satisfy the charge 
quantization condition:
\be g_m{\tilde g}_m = 4\pi n, \qquad n\in Z,\lb{101}\ee
if matter fields belong to the adjoint representation of $SU(N)$ group. The Dirac relation:
\be g_m{\tilde g}_m = 2\pi n, \qquad n\in Z,\lb{102}\ee
takes place for matter fields transforming according to fundamental representations (see 
Ref.~\ct{8,8i}). In general, we always have the following condition:
\be\alpha (\mu )\tilde \alpha (\mu) ={\mbox{const}},\lb{103}\ee
which is valid at arbitrary energies $\mu$.
\end{enumerate}

{\it We have no dual fundamental matter fields in the SM}. No experimental indications for any 
(abelian or non--abelian) fundamental monopoles or dyons. May be they exist at high energy 
scales.  

Considering QCD, we have quarks belonging to the triplet representation of $SU(3)$ colour gauge 
group, but light monopoles, belonging to the triplet representation of $\widetilde{SU(3)}$, are 
experimentally absent. By this reason, we have no dual symmetry for the total QCD. There exists 
only a dual symmetry of its gauge field part. The total QCD $\beta$--function is presented by 
curves 1', 2 and 3 of Fig.~\ref{f7}, instead of curves 1, 2, 3 describing by gluondynamics.

\section{Running  $\alpha_s^{-1}(\mu)$ }

Considering Eq.~(\ref{45}) for QCD regions 1', 2 and 3 of Fig.~\ref{f7}, we obtain the 
following results. 
\begin{enumerate}
\item[1.] For the region 1' we have:
\be \frac{\mbox{d}\ln \alpha(t)}{\mbox{dt}} = \beta(\alpha ),\lb{104}\ee
where $\beta(\alpha)$ ($\alpha\equiv \alpha_s$) is given by Eqs.~(\ref{50}) and (\ref{52}) with 
$N_f = 3,4,5$ up to the point $\mu = M_Z$. The solution of Eq.~(\ref{104}) in the 3--loop 
approximation has the following expression (see for example \ct{61,62}):
\be \alpha^{-1}(\mu) = \alpha^{-1}(\mu_R) + \frac{\beta_0}{4\pi}t + \lambda_1\ln
\frac{\alpha^{-1}(\mu)+\lambda_1}{\alpha^{-1}(\mu_R)+\lambda_1}\qquad {\mbox{for}} \qquad 
\alpha < 0.4,\lb{105}\ee
where $\mu_R$ is the renormalization point and:
\be \lambda_1 = \frac{1}{4\pi}\frac{\beta_1}{\beta_0}.\lb{106}\ee
\item[2.] For the region 2 we have:
\be\frac{\mbox{d}\ln\alpha(t)}{\mbox{dt}}=-\beta(\tilde\alpha)=-\beta(1/\alpha)\lb{107}\ee
where $\beta(\tilde \alpha)$ is given by Eqs.~(\ref{50}) and (\ref{51}). The solution of this 
equation gives the following behaviour:
\be\alpha^{-1}(\mu ) = \left[\alpha^{-1}(\mu_R) + \frac{\beta_0}{4\pi}t +\lambda_1\ln
\frac{\alpha^{-1}(\mu)+\lambda_1}{\alpha^{-1}(\mu_R)+\lambda_1}\right]^{-1}
\qquad {\mbox{for}} \qquad \alpha > 2.5,\lb{108}\ee
where $\beta_0$ and $\lambda_1$ correspond to $N_f=0$.
\item[3.] For the region 3 of Fig.~\ref{f7} we have Eq.~(\ref{83}), calculated according to the 
MAP--method. Curve 3 describes a sum of contributions of scalar monopoles given by 
$\beta_A(\alpha )$, and scalar ``monopoles" given by $-\beta_A(1/\alpha)$.
\end{enumerate}

The results of all these solutions (also valid for dyons) are presented by Fig.~\ref{f9} for 
the evolution of $\alpha_s^{-1}(\mu)$. The value $\mu_p\approx 1.2$ GeV, shown in 
Fig.~\ref{f9}, corresponds to the end of perturbation region 1' (or 1) and the beginning of 
confinement region 3. In the non--perturbation region 3, we have a solution given by solid 
curve 3 of Fig.~\ref{f9}, which approaches the point $\alpha_s(\mu )=1$ when $\mu\to 0$, and we 
see a rapid decrease of $\alpha^{-1}(\mu)$ near 1.  

It seems that solutions presented in Fig.~\ref{f9} by thin curves, which correspond to the 
region 2 and second part of the region 3 of Fig.~\ref{f7}, are not realized in QCD: they 
describe the running of inverse $\tilde \alpha$. We see that the dual part of QCD does not play 
an essential role in the formation of QCD vacuum: mainly monopoles, or magnetic part of dyons, 
participate in the formation of electric ``strings" -- ANO flux tubes, although solid curves 3 
of Fig.~\ref{f7} and Fig.~\ref{f9} present the contributions of both parts, electric and 
magnetic ones.

As it was shown in Refs.~\ct{63,64,65} (and developed in Refs.~\ct{61,62}), the QCD effective 
Lagrangian is given by the following expression:
\be L_{eff} = - \frac{\alpha_{eff}^{-1}(F^2)}{16\pi}F^2, \qquad {\mbox{where}}\qquad F^2 = 
F_{\mu\nu}^J F_{\mu\nu}^J\qquad (J=1,2,..8).\lb{109}\ee 
This Lagrangian contains the effective fine structure constant:
\be\alpha_{eff}(F^2) = \frac{g^2_{eff}}{4\pi},\lb{110}\ee 
which in general is a complicated nonlinear function of gluon fields. In the perturbative 
region $\alpha_{eff}(F^2)$ coincides with the running of $\alpha_{eff}(\mu)$, where 
$F^2=(\mu$ GeV)$^4$ (see Refs.~\ct{61,62,63,64,65}). But in the non--perturbative 
region we have a nontrivial situation: for asymptotically free theories the maximum of the 
effective action (e.g. minimum of the effective potential) already does not correspond to the 
classical vacuum with $F^2=0$. The quantum fluctuations lead to the formation of a gluon 
condensate (QCD vacuum). In our approach the gluon condensate 
$F_0^2=(\mu^*_{cond}$ GeV)$^4$ corresponds to the maximal value of $\alpha_s$ equal 
to 1. By this reason, we suppose that the variable $F^2$ is not given by $\mu^4$ in the 
non-perturbation region. Instead of $\mu $, it is natural to consider the following  variable:
\be\mu^* = \mu + \mu^*_{cond} \sim \frac 1r,\lb{111}\ee
which determines distances r, and we have:
\be F^2 = (\mu^* \,\,{\mbox{GeV}})^4.\lb{112}\ee
The nature of gluon condensate was investigated in a lot of papers (for example, very 
interesting considerations were given in Refs.~\ct{66,66i}).  

The value of gluon condensate was estimated in Refs.~\ct{67,67i,68}:
\be\left\langle\frac{\alpha_s}{4\pi}F^2_0\right\rangle
\approx 0.012\,\,{\mbox{GeV}^4}.\lb{113}\ee
For our case $\alpha_s=1$, and we have:
\be F^2_0\approx 0.15\,\,{\mbox{GeV}^4},\lb{114}\ee
or
\be\mu^*_{cond}\approx 0.62 \,\,{\mbox{GeV}}.\lb{115}\ee
The behaviour of inversed $\alpha(\mu)$ presented in Fig.~\ref{f9} shows that in the region 3 
given by solid curve we have: 
\be\alpha\simeq 0.45\pm 0.05,\lb{116}\ee 
e.g. almost unchanged (``freezing") $\alpha$ for a wide interval of $\mu^*$:
\be 0.72\,\,{\mbox{GeV}} \lesssim \mu^*\lesssim 1.82\,\,{\mbox{GeV}}.\lb{117}\ee  
We see that QCD, including its dual sector, acquires a new comprehension.

\section{The effective potential in QCD}

The perturbative effective potential is given by the following expression (see 
Refs.~\ct{61,62,63,64,65}):
\be V_{eff} = \frac{\alpha_{eff}^{-1}(F^2)}{16\pi}F^2
\qquad {\mbox{with}} \qquad F^2=(\mu\,\,{\mbox{GeV}})^4.\lb{118}\ee
However, this expression is not valid in the non--perturbative region, because the 
non--perturbative vacuum contains a condensation of chromo--magnetic flux tubes, according to 
so called ``spaghetti vacuum" by Nielsen--Olesen 
\ct{69}.
By this reason, we subtract the contribution of ``strings", determined by the gluon condensate 
$F_0^2$, from the expression (\ref{118}):
\be V_{eff} = \frac{\alpha_{eff}^{-1}(F^2)}{16\pi}F^2 - 
\frac{\alpha_{eff}^{-1}(F_0^2)}{16\pi}F_0^2,\lb{119}\ee 
using $F^2 = (\mu^*$ GeV)$^4$.

The behaviour of the effective potential $V_{eff}(F^2)$ is given by Fig.~\ref{f10}, and we see:
{\it The QCD effective potential shows a sharp minimum in the deep non--perturbative 
region} (at the point $F^2=F^2_0=0.15$ GeV$^4$). {\it This minimum points out the 
existence of the (unexpected) first order phase transition in QCD} at the point 
$\mu^*_{cond}\approx 0.62$ GeV.

\section{Conclusions}

In the present paper we have obtained the following results: 
\begin{enumerate}
\item[1.] The Zwanziger--type action can be constructed for nonabelian theories revealing the 
generalized dual symmetry. In the abelian limit this action corresponds to the Zwanziger 
formalism for quantum electro--magneto dynamics (QEMD). It was emphasized that although the 
generalized dual transformation is rather complicated, it is explicit in terms of loop space 
variables.
\item[2.] We have shown that the Zwanziger--type action confirms the invariance under the 
interchange:
$$\alpha \leftrightarrow \tilde \alpha = \frac {1}{\alpha}. $$
\item[3.] Such a symmetry leads to the generalized renormalization group
equations:
$$ \frac{\mbox{d}\ln \alpha(t)}{\mbox{dt}} = - \frac{\mbox{d}\ln {\tilde \alpha}(t)}{\mbox{dt}}
     = \beta(\alpha ) - \beta(\tilde \alpha) = \beta^{(total)}(\alpha),$$
where $\beta^{(total)}(\alpha)$ is the total $\beta$--function, 
{\it antisymmetric} under the interchange:
$$\alpha \leftrightarrow \tilde \alpha,\qquad {\mbox{or}} \qquad\alpha \leftrightarrow 
\frac 1{\alpha}$$
for pure nonabelian theories.
\item[4.] We have applied the method of Maximal Abelian Projection (MAP) by t' Hooft to the 
pure $SU(3)$ gauge theory with aim to describe the behaviour of the total $\beta$--function 
in the region $0 \le \alpha, \tilde \alpha < \infty$.
\item[5.] We have shown that as a result of the dual symmetry and MAP $\beta^{(total)}(\alpha)$ 
has a zero at $\alpha = \tilde \alpha = 1$ ({\it``fixed point"}): 
$$\beta^{(total)}(\alpha=\tilde \alpha = 1) = 0.$$
\item[6.] At the first step, we have considered the existence of the $N-1$ Higgs abelian scalar 
monopole fields ${\tilde \phi}_{1,2,..., N-1}$ and $N-1$ Higgs abelian scalar electric fields 
$\phi_{1,2,..., N-1}$ in the non--perturbative region of pure nonabelian 
$SU(N)\times \widetilde{SU(N)}$ gauge theories.
\item[7.] At the second step, we have assumed that a generalized dual symmetry naturally leads 
to the existence of the Higgs scalar dyon fields $\phi_{1,2,..., N-1}$, which are created by 
non--perturbative effects of the $SU(N)\times \widetilde{SU(N)}$ gluondynamics. These abelian 
dyons have both (electric and magnetic) charges, and describe the total $\beta$--function in 
the following non--perturbative region:
$$ 0.4 < \alpha < 2.5,$$
which explains the freezing of $\alpha_s$ in QCD.
\item[8.] We also discussed the case of nonabelian theories with matter fields, which in 
general have no dual symmetry.
\item[9.] We have investigated the running of $\alpha_s^{-1}(\mu)$ in the perturbative and 
non--perturbative regions of $\mu$.
\item[10.] We have calculated the value of the gluon condensate: 
$F_0^2\approx 0.15$ GeV$^4$, in accord with the well--known result, given by 
literature.
\item[11.] We have presented the behaviour of the QCD effective potential as a function of 
$F^2$, having a sharp minimum in the non--perturbative region. This minimum, corresponding to 
the gluon condensate, prompts the existence of the first order phase transition in QCD. 
\end{enumerate}

\section*{Acknowledgements:}

One of the authors (L.V.L.) deeply thanks the Niels Bohr Institute, where this investigation 
was born, and the Institute of Mathematical Sciences (Chennai, India), personally 
Prof.~N.D.~Hari Dass, for hospitality, financial support and useful discussions. We are 
thankful of the Organizing Committee of the 12th Lomonosov Conference on Elementary Particle 
Physics (Moscow, Russia), where our talk \ct{70} was presented in August 2005.
 
This work was supported by the Russian Foundation for Basic Research (RFBR), project $N^o$ 
05--02--17642, and we thank RFBR.

\section*{Appendix A: The regularization procedure}
\setcounter{equation}{0}
\renewcommand{\theequation}{A.\arabic{equation}}

With aim to understand the difference between the quantities $F_{\mu}[\xi|s]$ and 
$E_{\mu}[\xi|s]$, it is convenient to give some explanations, coming from 
Refs.~\ct{9,9i,9ii,10,10i,11}.
The new variables $E_{\mu}[\xi|s]$ are not gauge invariant like $F_{\mu}[\xi|s]$. But in spite 
of this inconvenient property, the variables $E_{\mu}[\xi|s]$ are more useful for studying the 
generalized duality. They can be represented as the bold curve in Fig.~\ref{f4} where the phase 
factors $\Phi_{\xi}(s,0)$ cancel parts of the faint curve representing $F_{\mu}[\xi|s]$. The 
loop derivative considered in this paper is defined as
\be\frac{\delta \Phi[\xi ]}{\delta \xi^{\mu}(s)} = \lim_{\Delta \to 0}
\frac {\Phi [\xi'] - \Phi [\xi]}{\Delta}, \lb{a1}\ee
where
\be {\xi'}^{\lambda} = {\xi}^{\lambda}(s') +
\Delta \delta^{\lambda}_{\mu}\delta (s - s').\lb{a2}\ee
The $\delta$--function $\delta (s-s')$ is a bump function centred at s with width $\epsilon = 
s_{+} - s_{-}$ (see Fig.~\ref{f4}). In contrast to $F_{\mu}[\xi|s]$, the quantity 
$E_{\mu}[\xi|s]$ depends only on a ``segment" of the loop $\xi^{\mu}(s)$ from $s_{-}$ to 
$s_{+}$. The regularization of $\delta$--function is necessary for the definition of loop 
derivatives used in this theory. The quantities $E_{\mu}[\xi|s]$, constrained by the condition:
\be\frac{\delta E_{\mu}[\xi|s]}{\delta \xi^{\nu}} -
\frac{\delta E_{\nu}[\xi|s]}{\delta \xi^{\mu}} = 0,\lb{a3}\ee
constitute a set of the curl--free variables valid for the description of nonabelian theories 
revealing  properties of the generalized dual symmetry.

\section*{Appendix B: Renormalization group equation for non-dual
and dual coupling constants}
\setcounter{equation}{0}
\renewcommand{\theequation}{B.\arabic{equation}}

The renormalization group (RG) describes an independence of a theory and its 
couplings on an arbitrary scale parameter $M$. We are interested in RG applied 
to the effective potential depending on scalar field $\phi$. 
The renormalization group equation (RGE) for the 
effective potential means that the potential 
cannot depend on a change in the arbitrary renormalization scale parameter $M$:
\be\frac{dV_{eff}}{dM}.\lb{b0}\ee
The effects of changing it are absorbed into changes in the coupling constants,
 masses and fields, giving so--called running quantities. Knowing the 
dependence on $M^2$ is equivalent to 
knowing the dependence on $\phi^2$. This dependence is given by 
RGE. Considering 
the RGE improvement of the potential, we follow the approach by Coleman and 
Weinberg \ct{71} (see also the review \ct{72}) for scalar electrodynamics
and its extension to the massive theory \ct{73}. Here we have the difference 
between the scalar electrodynamics \ct{71} and scalar 
QuantumElectroMagnetoDynamics (QEMD) when we have scalar particles 
with electric charge $e$ and scalars with magnetic charge $g$.

RGE for the improved one--loop effective potential can be given in QEMD by the 
following expression:
\be\left(M\frac{\partial}{\partial M} +\beta_{\lambda}\frac{\partial}{\partial 
\lambda} +
e\beta_e\frac{\partial}{\partial e} + g\beta_g\frac{\partial}{\partial g} +
\beta_{(m^2)}{m^2}\frac{\partial}{\partial m^2} - \gamma \phi 
\frac{\partial}{\partial \phi}\right) V_{eff}(\phi^2) = 0, \lb{b1}\ee
where the function $\gamma $ is the anomalous dimension:
\be \gamma\left(\frac{\phi}M\right) = - \frac{\partial \phi}{\partial M}. 
                      \lb{b2}\ee
RGE (\ref{b1}) leads to a new improved effective potential \ct{71}:
\be V_{eff}(\phi^2) = \frac 12 m^2_{ren}(t) G^2(t)\phi^2 +
\frac 14 \lambda_{ren}(t) G^4(t) \phi^4, \lb{b3}\ee
where
\be G(t)\equiv \exp\left[ - \frac 12 \int_0^t dt' \gamma(g_{ren}(t'),
\lambda_{ren}(t'))\right]. \lb{b4}\ee
Eq.~(\ref{b1}) reproduces also a set of ordinary differential equations:
\be\frac{d\lambda_{ren}}{dt} = \beta_{\lambda}(g_{ren}(t),\lambda_{ren}(t)), 
\lb{b5}\ee
\be\frac{dm^2_{ren}}{dt} = m^2_{ren}(t) \beta_{(m^2)}(g_{ren}(t),\lambda_{ren}
(t)), \lb{b6}\ee
\be\frac{d\ln e_{ren}}{dt} = - \frac{d\ln g_{ren}}{dt} = \beta_e(g_{ren}(t),
\lambda_{ren}(t)) - \beta_g (g_{ren}(t),\lambda_{ren}(t))\equiv \beta^
{(total)}, \lb{b7}\ee
where $t = \ln (\phi^2/{M^2})$, and the subscript ``ren" means the 
``renormalized" quantity.

The last equation (\ref{b7}) is obtained with the help of the Dirac relation 
$eg=2\pi n$ $(n\in Z)$ for minimal charges when $eg=2\pi$. 
Indeed, in Eq.~(\ref{b1}):
$$ e\beta_e\frac{\partial}{\partial e} + g\beta_g\frac{\partial}{\partial g} =  
e\beta_e\frac{\partial}{\partial e} + g\beta_g\frac{de}{dg}\frac{\partial}
{\partial e} 
= \left(e\beta_e + g\beta_g\left(-\frac{2\pi}{g^2}\right)\right)\frac{\partial}
{\partial e} $$ 
\be = (\beta_e - \beta_g)\frac{\partial}{\partial \ln e} 
   = \beta^{(total)}\frac{\partial}{\partial \ln e},   \lb{b8} \ee
where
\be\beta^{(total)} = \beta_e - \beta_g.  \lb{b9}\ee
Using fine structure constants
\be\alpha = \frac{e^2}{4\pi},\qquad\tilde \alpha = \frac{g^2}{4\pi}, 
\lb{b10}\ee 
we obtain Eq.(\ref{45}) given in Section 5. But for abelian theories we
have $\alpha\tilde \alpha=1/4$.

Eq.(\ref{45}) takes place also for nonabelian theories with charge 
quantization condition $g\tilde g=4\pi n$ $(n\in Z)$ giving
$\alpha\tilde \alpha=1$.

\section*{Appendix C: Abelian projection method}
\setcounter{equation}{0}
\renewcommand{\theequation}{C.\arabic{equation}}

In this Appendix we present the method of the Maximal Abelian Projection (MAP) by G. t'~Hooft 
\ct{31}, following the review \ct{33}.

For any composite field $X$ (for example, $F_{\mu\nu}$) transforming as an adjoint 
representation 
\be X\to X'=VXV^{-1},\lb{c1}\ee
we can find  the specific unitary matrix $V$ (the gauge), where $X$ is diagonal:
\be X'=VXV^{-1}=diag (\lambda_1,\lambda_2,...\lambda_N).\lb{c2}\ee
For $X$ from the Lie  algebra of  $SU(N)$, one can choose 
$\lambda_1\leq\lambda_2\leq\lambda_3\leq...\lambda_N$. It is clear that $V$ is determined up to 
the left multiplication by a diagonal $SU(N)$, which belongs to the Cartan (or largest abelian)
subgroup of $SU(N)$: 
\be U(1)^{N-1}\subset SU(N).\lb{c3}\ee

Now we transform $A_{\mu}$, according to the gauge (\ref{c2}):
\be\tilde A_{\mu}=V\left(A_{\mu}+\frac{i}{g}\partial_{\mu}\right)V^{-1} \lb{c4}\ee
and consider the transformations of $\tilde A_{\mu}$ under $U(1)^{N-1}$. The diagonal 
components 
\be a^i_{\mu}\equiv (\tilde A_{\mu})_i^i   \qquad (i=1,2,3)\lb{c5}\ee
transform as ``photons'':
\be a_{\mu}^i\to a^i_{\mu}=a^i_{\mu}+\frac{1}{g}\partial_{\mu}\alpha_i,\lb{c6}\ee
while nondiagonal, $c_{\mu}^{ij}  \equiv A_{\mu}^{ij}$, transform as charged fields:
\be C^{'ij}_{\mu}= \exp [i(\alpha_i-\alpha_j)] C_{\mu}^{ij}.\lb{c7}\ee
By 't Hooft remarks \ct{31}, this is not the whole story: there appear singularities due to a 
possible  coincidence of two or more eigenvalues $\lambda_i$, which leads to the existence of
monopoles. To make it explicit, let us consider (as in \ct{31}) the ``photon" field strength:
$$ f^i_{\mu\nu}=\partial_{\mu}
a^i_{\nu}-\partial_{\nu}a^i_{\mu}$$
\be =VF_{\mu\nu}V^{-1}+ig\left[V\left(A_{\mu}+\frac{i}{g}\partial_{\mu}\right)V^{-1},
V\left(A_{\nu}+\frac{i}{g}\partial_{\nu}\right)V^{-1}\right],      \lb{c8}\ee
and define the monopole current:
\be K^i_{\mu}=\frac{1}{8\pi}\varepsilon_{\mu\nu\rho\sigma}\partial_{\nu}
 f^i_{\rho\sigma},\qquad\partial_{\mu}K^i_{\mu}=0.\lb{c9}\ee
Since $F_{\mu\nu}$ is regular, the only singularity giving rise to $K^i_{\mu}$ is the 
commutator term in (\ref{c8}), otherwise the smooth part of $a^i_{\mu}$ does not contribute to 
$K^i_{\mu}$ because of  the antisymmetric tensor.

Hence one can define the magnetic charge $m^i(\Omega)$ in the $3d$ region $\Omega$:
\be m^i(\Omega)=\int_{\Omega}d^3\sigma_{\mu}K^i_{\mu}= \frac{1}{8\pi}\int_{\partial\Omega}
 d^2\sigma_{\mu\nu}\varepsilon_{\mu\nu\rho\sigma}f^i_{\rho\sigma}.\lb{c10}\ee
Let us consider now the situation when two eigenvalues of (\ref{c2}) coincide, e.g. 
$\lambda_1=\lambda_2$. This may happen at one $3d$ point in $\Omega$, $x^{(1)}$, 
i.e. on the line 
in the $4d$, which one can visualize   as the magnetic monopole world line. The contribution to
$m^i(\Omega)$ comes  only from the infinitesimal neighbourhood $B_{\varepsilon}$ of $x^{(1)}$:
\bea m^i(B_{\varepsilon}(x^{(1)}))&=&\frac{i}{4\pi}\int_{S_{\varepsilon}}
d^2\sigma_{\mu\nu}\varepsilon_{\mu\nu\rho\sigma} \left[V\partial_{\rho}V^{-1},
V\partial_{\sigma}V^{-1}\right]_{ii}\nonumber\\
&=&-\frac{i}{4\pi}\int d^2\sigma_{\mu\nu}\varepsilon_{\mu\nu\rho\sigma}\partial_{\rho}
\left[V\partial_{\sigma}V^{_1}\right]_{ii}.\lb{c11}\eea
The  term $V\partial_{\sigma}V^{-1}$ is singular and should be treated with care. To make it 
explicit, one can write:
\be V=W\left(\begin{array}{cc}\cos\frac{1}{2}\theta+i\vec{\sigma}\vec e_{\phi}
\sin\frac{\theta}{2}& 0\\0&1\end{array}\right),\lb{c12}\ee
where $W$ is a smooth $SU(N)$ function near $x^{(1)}$. Inserting it in (\ref{c11}), one obtains:
\be m^i(B_{\varepsilon}(x^{(1)}))=\frac{1}{8\pi}\int_{S_{\varepsilon}}d^2\sigma_{\mu\nu}
\varepsilon_{\mu\nu\rho\sigma}\partial_{\rho}(1-\cos \theta)\partial_{\sigma}
\phi[\sigma_3]_{ii},\lb{c13}\ee
where $\phi$ and $\theta$ are azimuthal and polar angles. The integrand in (\ref{c13}) is a 
Jacobian displaying a mapping from $S^2_{\varepsilon}(x^{(1)})$ to $(\theta,\phi)\sim 
SU(2)/U(1)$. Since
\be\Pi_2\left(\frac{SU(2)}{U(1)}\right)=Z,\lb{c14}\ee
the magnetic charge is $m^i=0, \pm 1/2, \pm 1,..$.

From the derivation above it is clear, that the point $x=x^{(1)}$, where $\lambda_1(x^{(1)})=
\lambda_2(x^{(1)}),$ is a singular point of the gauge transformed $A_{\mu}$ and $a^i_{\mu}$, 
and the latter behaves near $x=x^{(1)}$ as $0(1/|x-x^{(1)}|)$, and abelian projected field 
strength $f^i_{\mu\nu}$ is $0(1/|x-x^{(1)}|^2)$, similar to the magnetic field of a point--like 
magnetic monopole. However, several properties should be stressed now:

\begin{enumerate}
\item[1.] The original vector potential $A_{\mu}$ and $F_{\mu\nu}$  are smooth and do not show 
any singular behaviour.
\item[2.] At large distances $f^i_{\mu\nu}$ is not, generally speaking, monopole--like, i.e. 
does not decrease as $1/|x-x^{(1)}|^2$, so that similarity to the magnetic monopole (its 
topological properties) can be seen only in the vicinity of the singular point $x^{(1)}$.
\item[3.] In general, monopoles considered by MAP--method have nothing to do with classical 
solutions: MAP--monopoles are quantum fluctuations of gluon fields. Actually almost any field 
distribution in the vacuum may be abelian projected into $a_{\mu}^i,f_{\mu\nu}^i$  and then
magnetic monopoles can be detected.
\end{enumerate}

\clearpage\newpage
\bfi
\centering
\includegraphics[height=110mm,keepaspectratio=true]{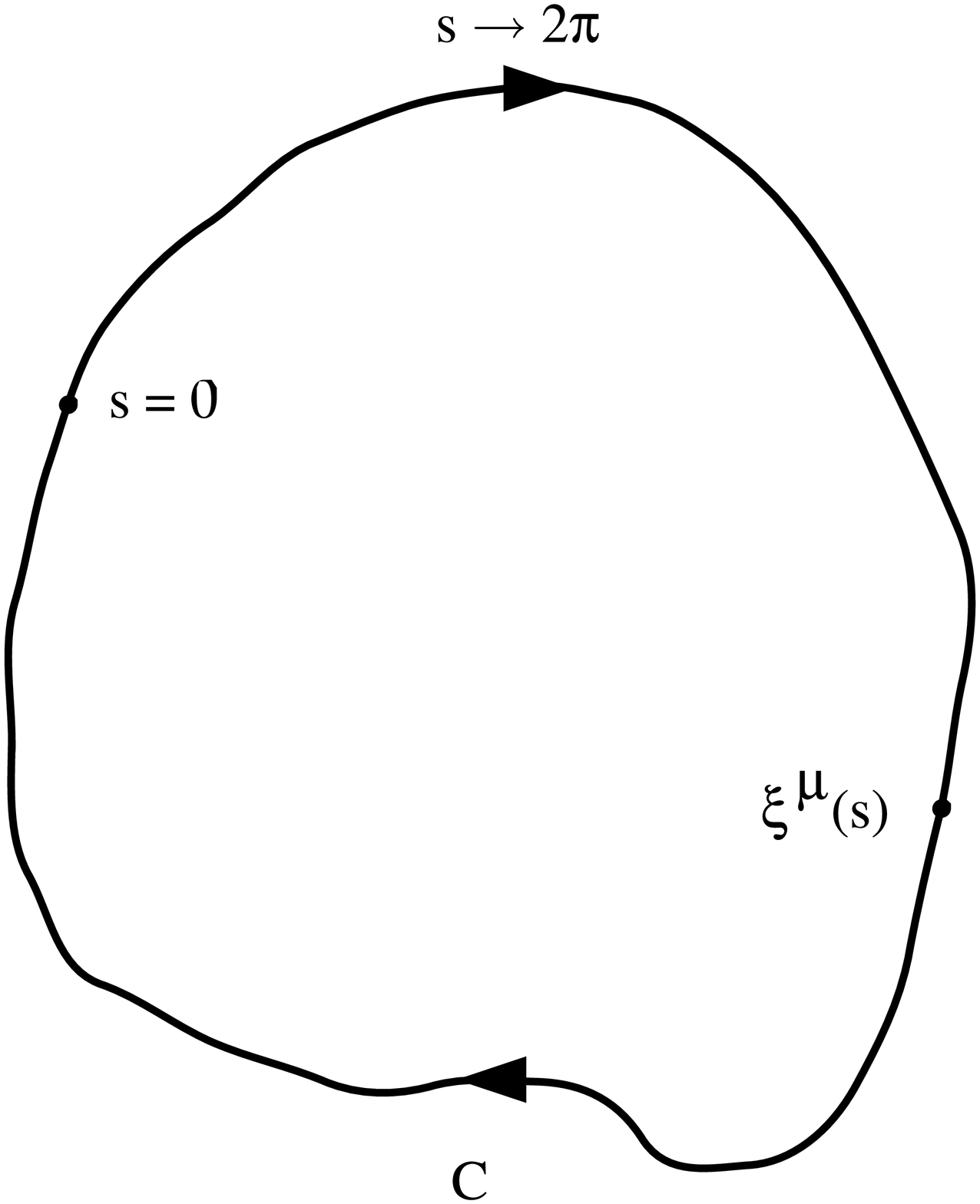}
\caption{The closed loop $C$ with coordinates $\xi^{\mu}(s)$ and loop parameter s: 
$0\le s\le 2\pi$.}
\lb{f1}
\efi

\clearpage\newpage
\bfi
\centering
\includegraphics[height=110mm,keepaspectratio=true]{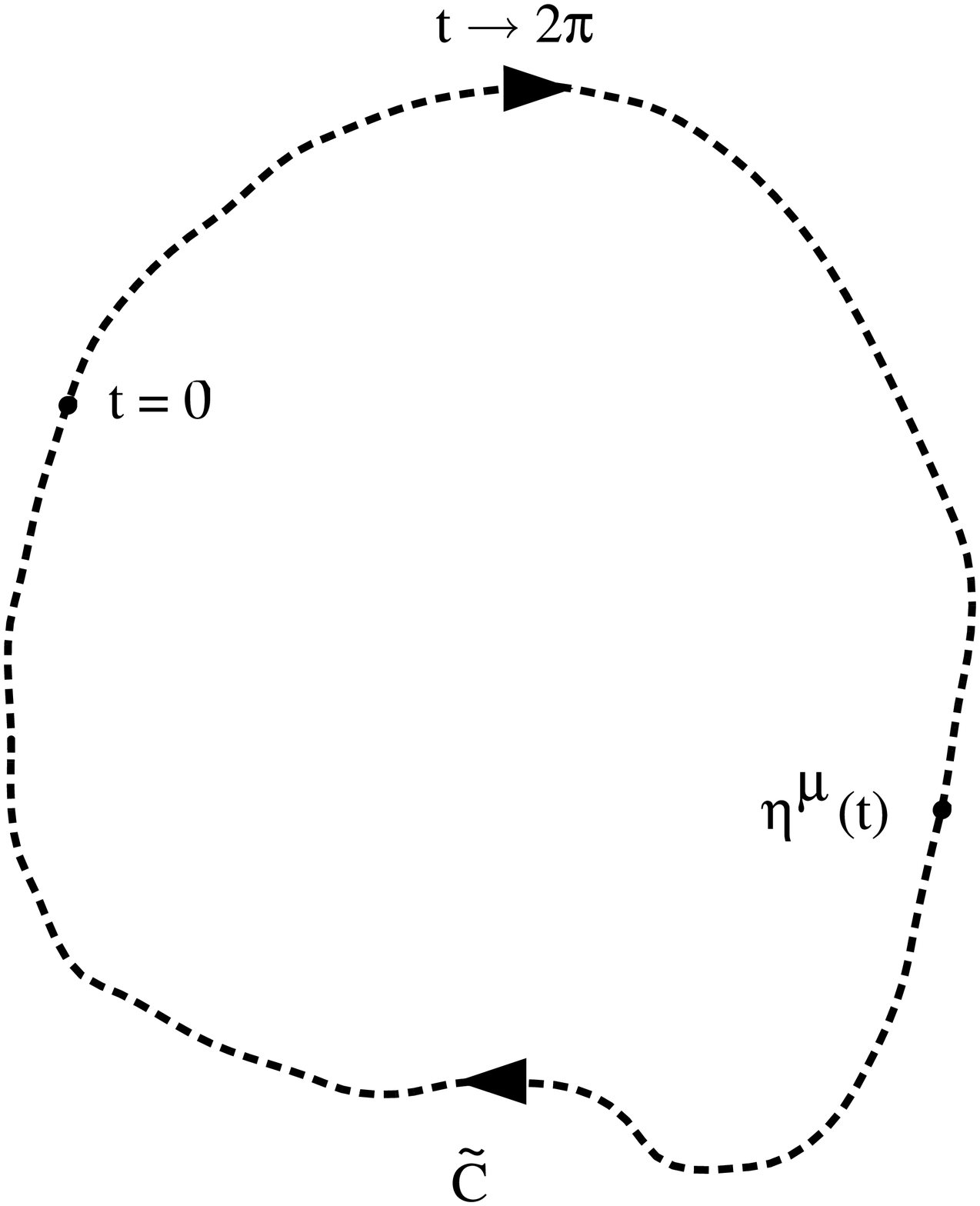}
\caption{The closed loop $\tilde C$ with coordinates $\eta^{\mu}(t)$ and loop parameter t: 
$0\le t\le 2\pi$.}
\lb{f2}
\efi

\clearpage\newpage
\bfi
\centering
\includegraphics[height=120mm,keepaspectratio=true]{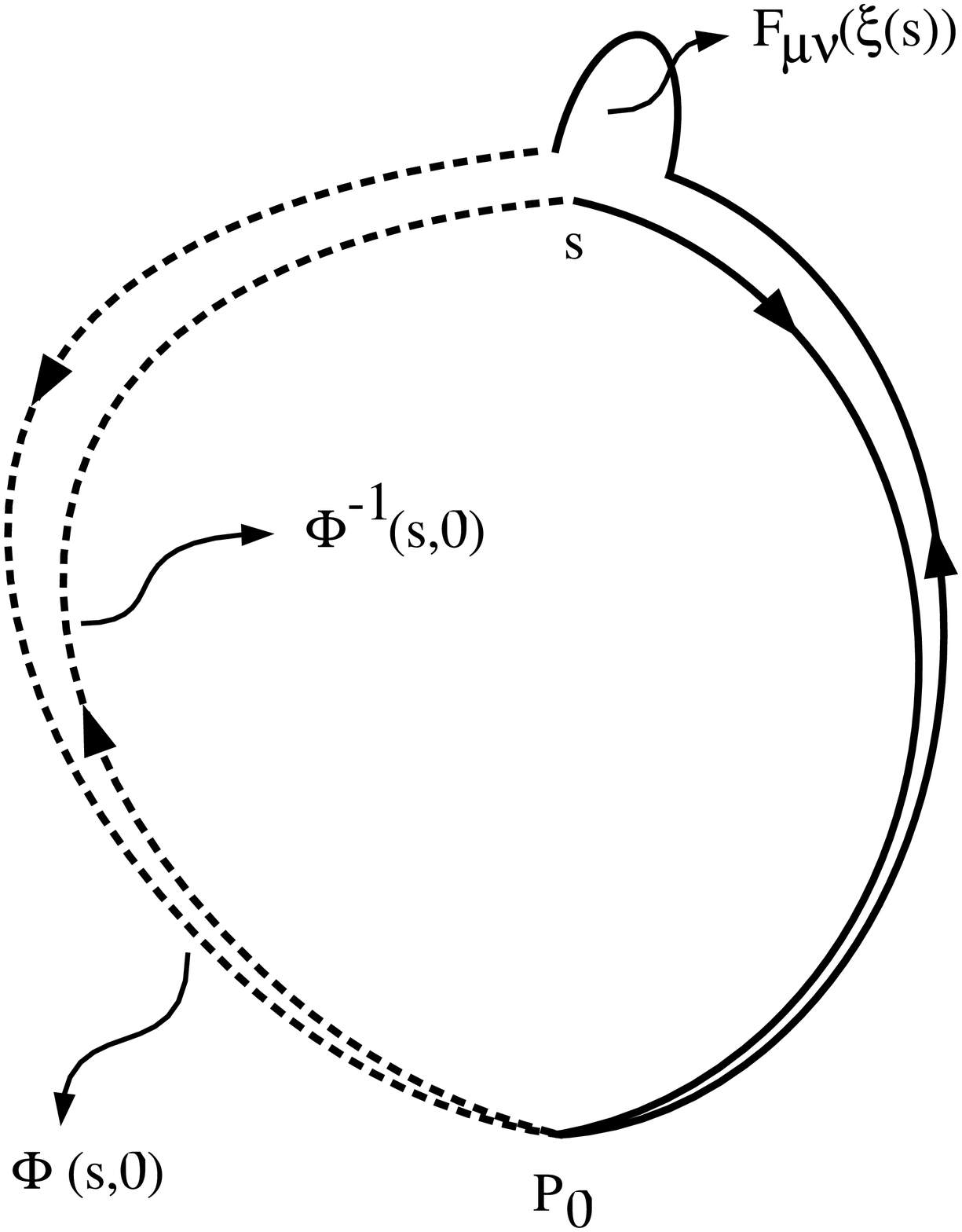}
\caption{The illustration of the quantity $F_{\mu}[\xi|s]$.}
\lb{f3}
\efi

\clearpage\newpage
\bfi
\centering
\includegraphics[height=100mm,keepaspectratio=true]{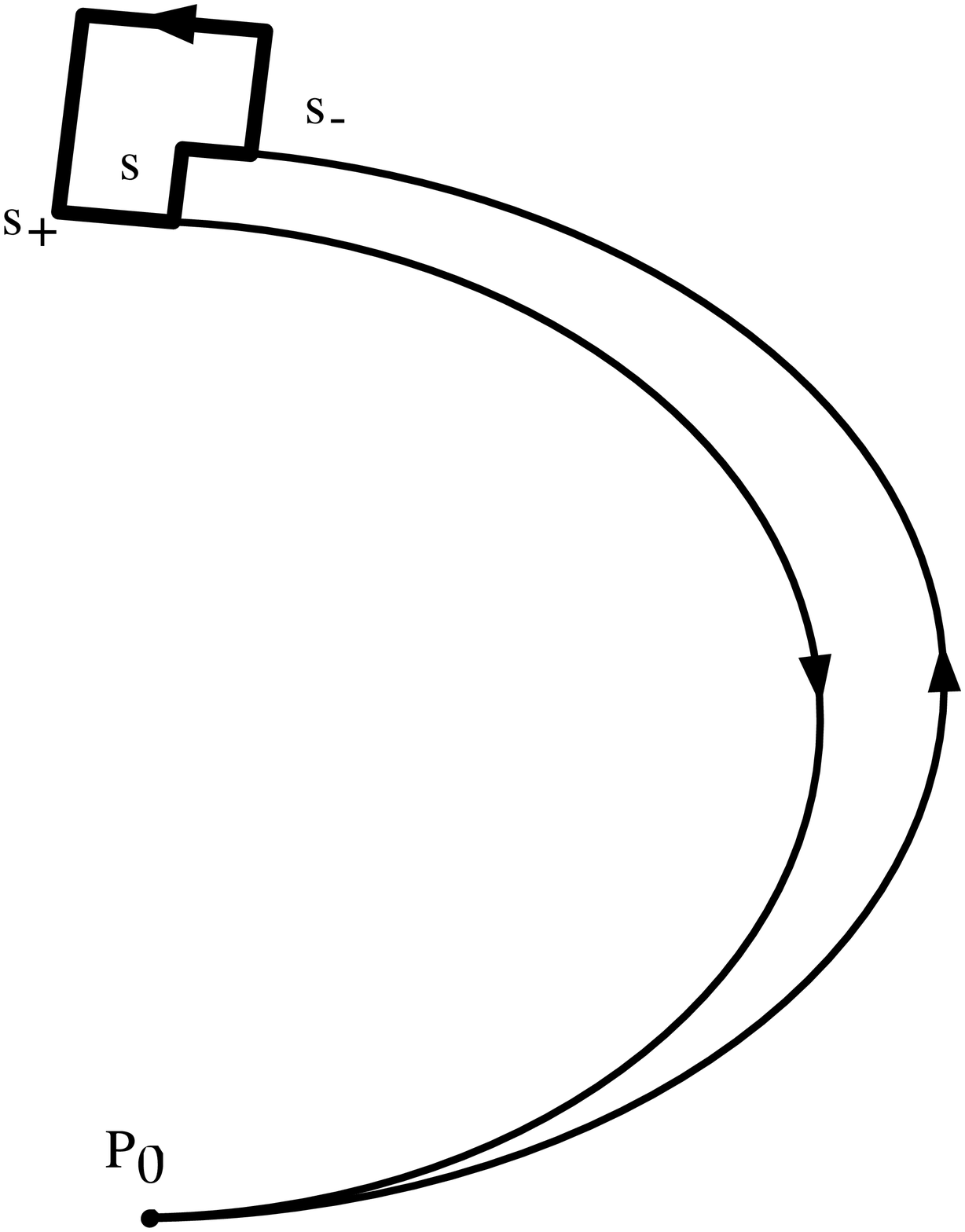}
\caption{The illustration of the quantity $E_{\mu}[\xi|s]$.}
\lb{f4}
\efi

\clearpage\newpage
\bfi
\centering
\includegraphics[height=165mm,keepaspectratio=true,angle=-90]{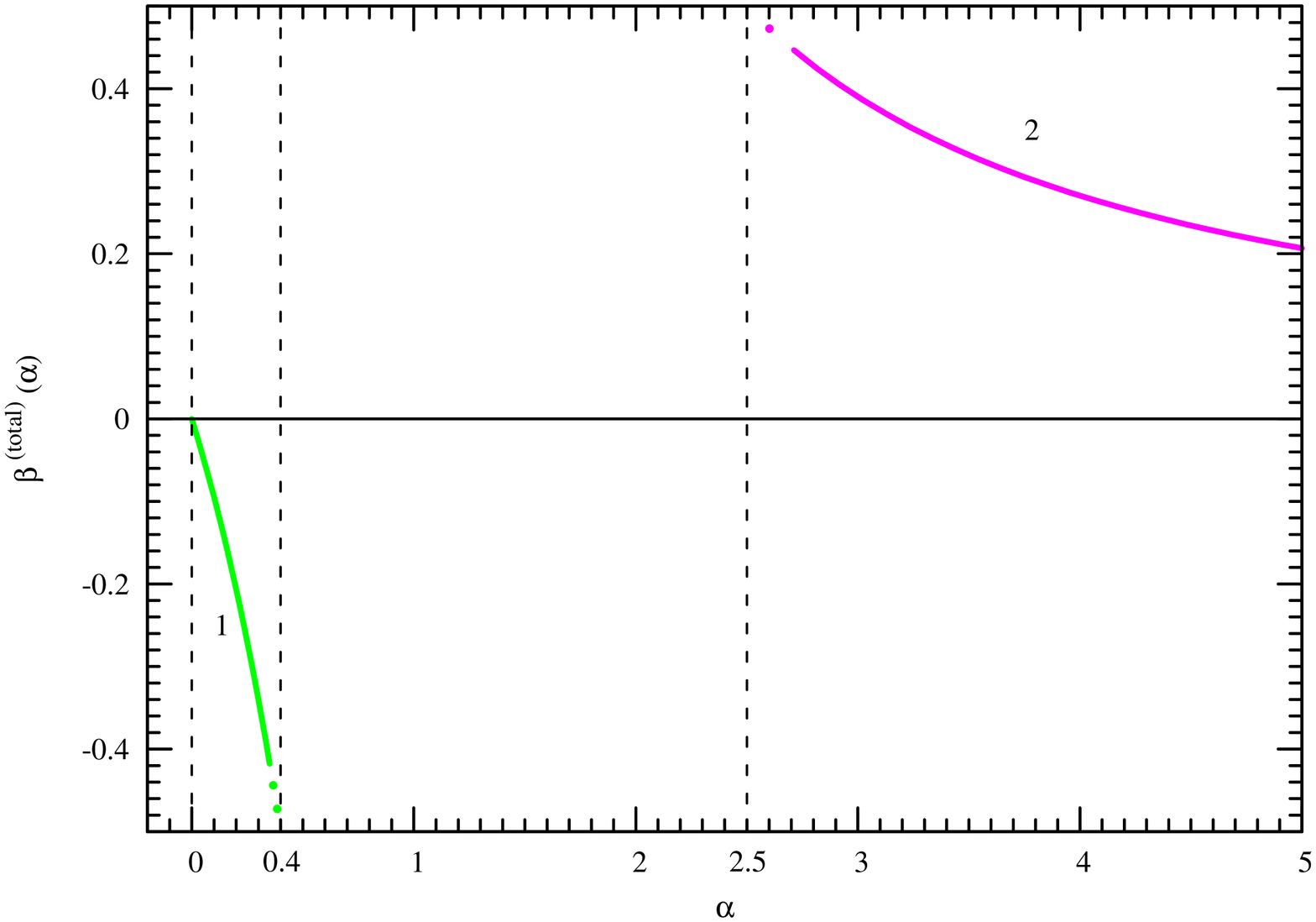}
\caption{The perturbative $\beta$--functions of the pure $SU(3)\times \widetilde{SU(3)}$ color 
gauge theory: $\beta(\alpha)$ -- curve 1, and $-\beta(\tilde \alpha)= -\beta(1/\alpha)$ -- 
curve 2.}
\lb{f5}
\efi 

\clearpage\newpage
\bfi
\centering
\includegraphics[height=70mm,keepaspectratio=true]{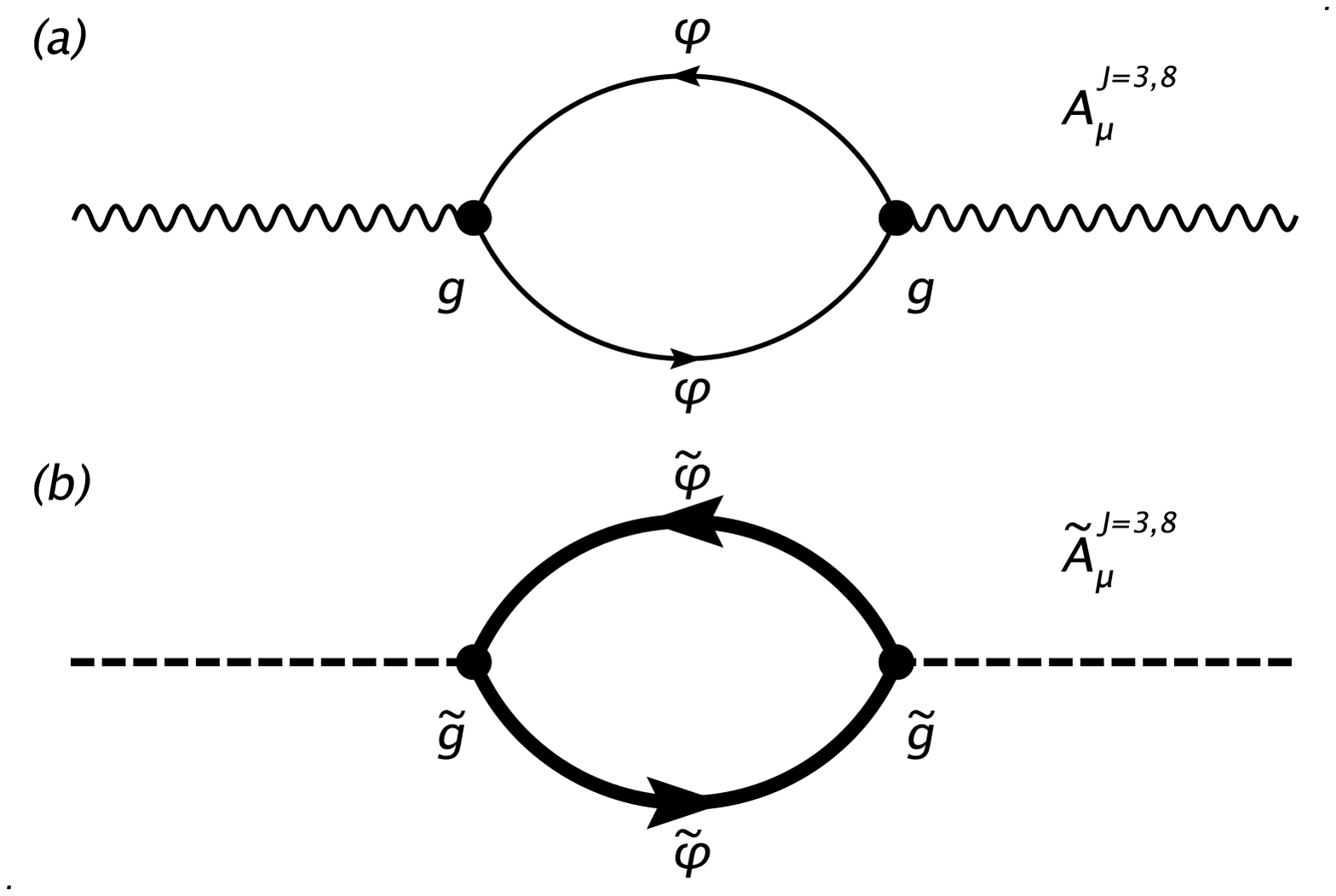}
\caption{The one--loop corrections: (a) from electric ``monopoles'' to the gluon propagator, 
and (b) from monopoles to the dual gluon propagator.}
\lb{f6}
\efi

\clearpage\newpage
\bfi
\centering
\includegraphics[height=165mm,keepaspectratio=true,angle=-90]{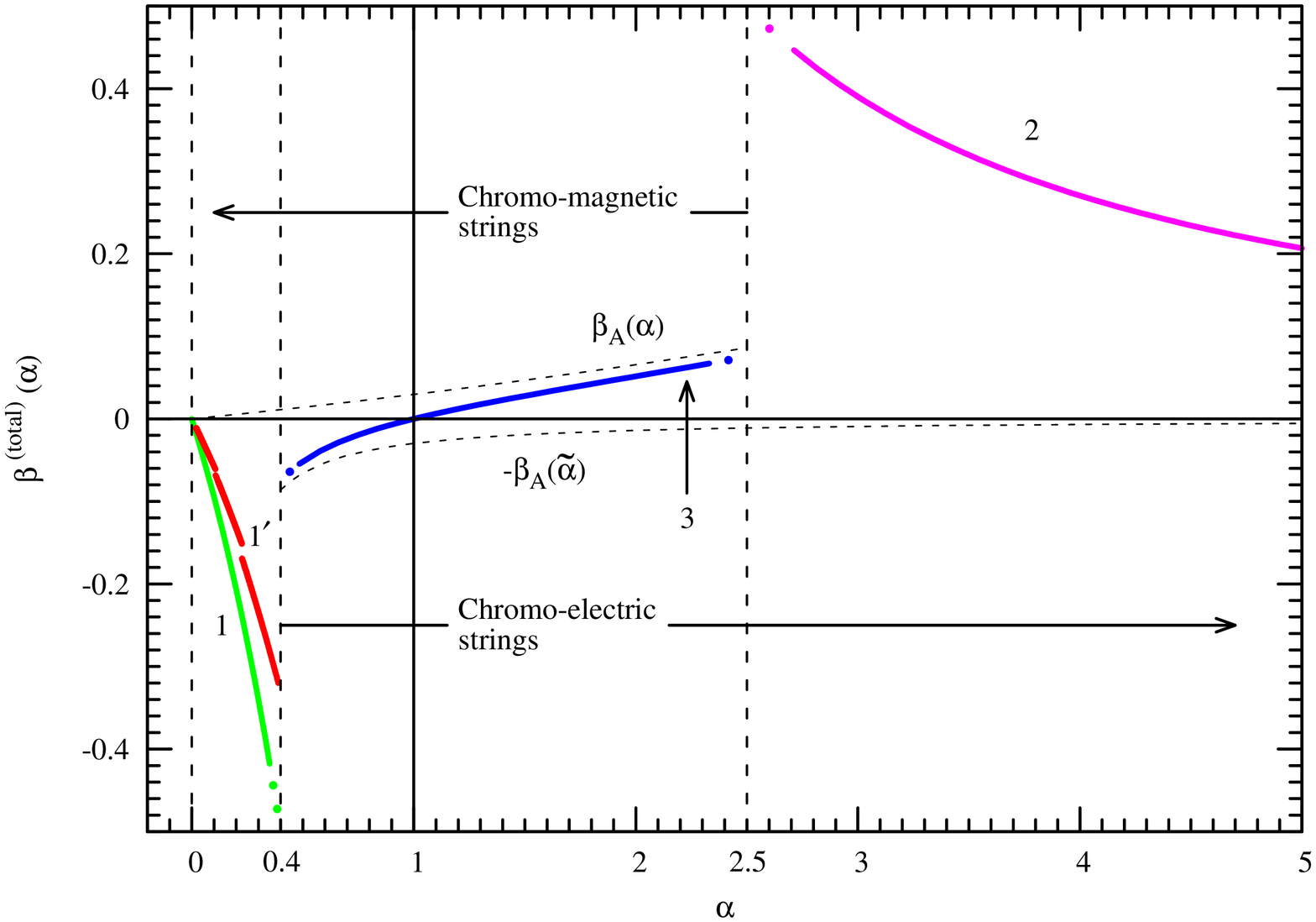}
\caption{The total $\beta$--function for QCD and pure $SU(3)\times \widetilde{SU(3)}$ color 
gauge theory (gluondynamics): (i) Curve 1 describes the perturbative $\beta$--function, 
corresponding to contributions of usual gluons, presented in (a) of Fig.~\ref{f8}. (ii) Curve 
1' corresponds to the perturbative QCD $\beta$--function (with quark and gluon contributions).
(iii) Curve 2 describes the perturbative $\beta$--function, corresponding to contributions of 
dual gluons, presented by (b) of Fig.~\ref{f8}. (iv) Curve 3 describes a sum of contributions 
of scalar ``monopoles'', given by $\beta_A(\alpha )$, and scalar monopoles, given by 
$-\beta_A(1/\alpha)$. Both of them exist in the non--perturbative  region of gluondynamics, or 
QCD. The critical points: $\alpha_1\approx 0.4$ and $\alpha_2\approx 2.5$, and regions of the 
existence of chromo--electric (for $\alpha\ge 0.4$) and chromo--magnetic (for $\alpha\le 2.5$)
strings (ANO flux--tubes) are also shown in this figure. The total $\beta$--function has a zero 
at the point $\alpha =\tilde \alpha=1$, predicted by our model.}
\lb{f7}
\efi

\clearpage\newpage
\bfi
\centering
\includegraphics[height=105mm,keepaspectratio=true]{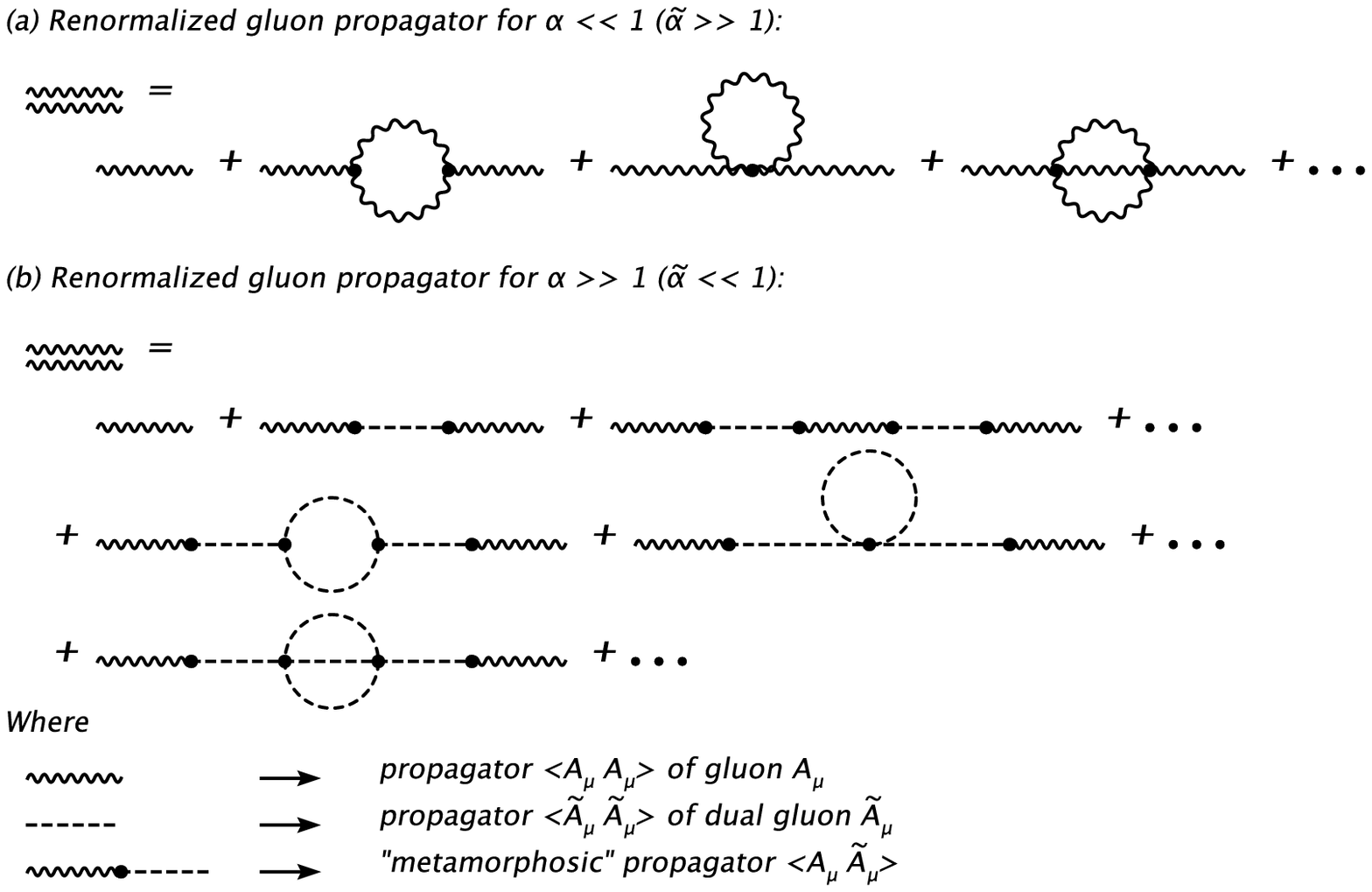}
\caption{The loop contributions to the gluon propagator (without matter fields): (a) the 
contributions from usual gluons, and (b) the contributions from dual gluons.}
\lb{f8}
\efi

\clearpage\newpage
\bfi
\centering
\includegraphics[height=165mm,keepaspectratio=true,angle=-90]{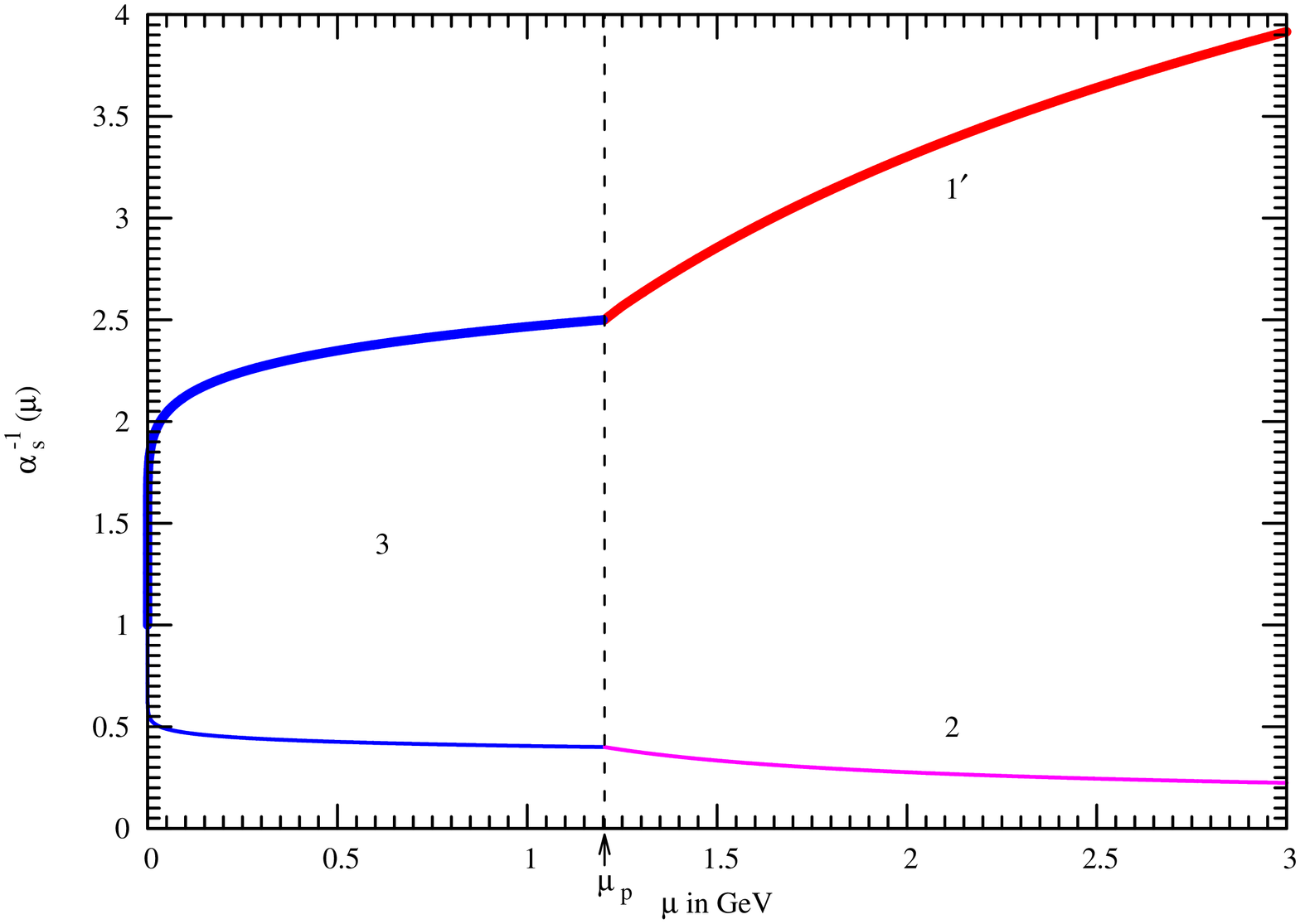}
\caption{The evolution of the inversed $\alpha(\mu)\equiv \alpha_s(\mu)$. A solid curve 
corresponds to the curve 1' and first part of region 3 of Fig.~\ref{f7}. A thin curve 
corresponds to the region 2 and second part of the same region 3 of Fig.~\ref{f7}. The 
solutions, presented by this curve, are not realized in QCD: they describe the evolution of the 
inversed $\tilde \alpha$. In the non--perturbative region, the solid curve 3 approaches to the
point $\alpha(0)=1$. We see the sharp decreasing of $\alpha^{-1}(\mu)$ near this point.}
\lb{f9}
\efi

\clearpage\newpage
\bfi
\centering
\includegraphics[height=165mm,keepaspectratio=true,angle=-90]{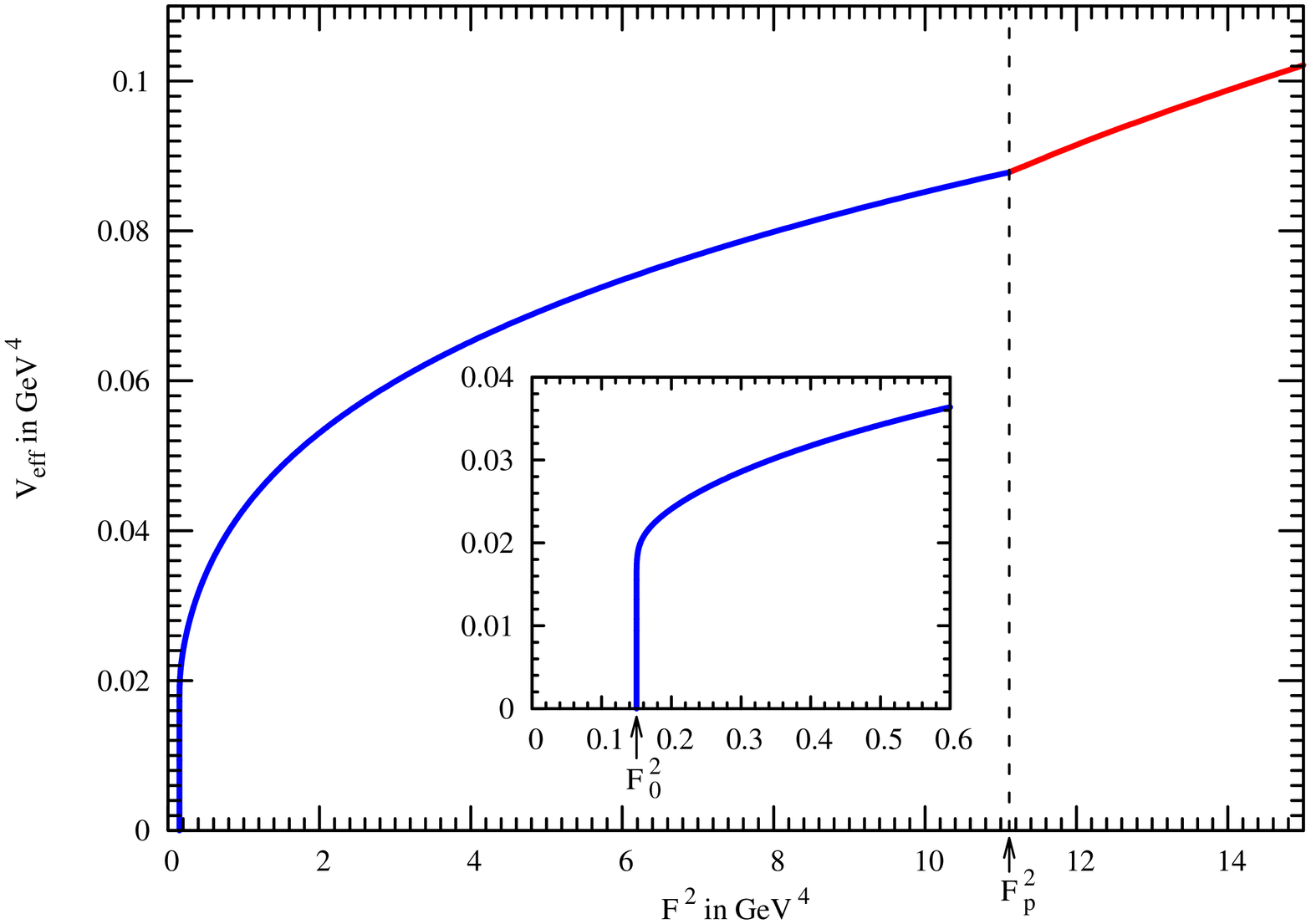}
\caption{The effective potential of $SU(3)\times \widetilde{SU(3)}$ color gauge theory, as a 
function of $F^2$. The point $F^2=F_0^2\approx 0.15$ GeV$^4$ corresponds to the gluon 
condensate. Near this point the effective potential has a sharp minimum, corresponding to the 
first order phase transition in the deep non--perturbative region, far from its beginning 
$F_p^2$.}
\lb{f10}
\efi


\begin{thebibliography}{99}
\bibitem{1}
S.~Mandelstam, Phys.Rev. {\bf D11}, 3026 (1975).
\bibitem{2}
G. 't~Hooft, Nucl.Phys. {\bf B138}, 1 (1978).
\bibitem{3}
J.~Maldacena, Adv.Theor.Math.Phys. {\bf 2}, 231 (1998); arXiv: hep-th/9711200.
\bibitem{4}
M.B.~Halpern, Phys.Rev. {\bf D16}, 1798 (1977).
\bibitem{4i}
M.B.~Halpern, Phys.Rev. {\bf D16}, 3515 (1977).
\bibitem{4ii}
M.B.~Halpern, Phys.Rev. {\bf D19}, 517 (1979). 
\bibitem{4iii}
M.B.~Halpern, Nucl.Phys. {\bf B139}, 477 (1978).
\bibitem{5}
G.G.~Batrouni, Nucl.Phys. {\bf B208}, 12 (1982).
\bibitem{5i}
G.G.~Batrouni, Nucl.Phys. {\bf B208}, 467 (1982).
\bibitem{6}
G.G.~Batrouni and M.B.~Halpern, Phys.Rev. {\bf D30}, 1782 (1984).
\bibitem{7}
H.M.~Chan and S.T.~Tsou, 
{\it An Electric--Magnetic Duality for Non--Abelian Yang--Mills Fields}, The 
28th Int.Conf. on High--energy Physics, Warsaw 1996, ICHEP'96, {\bf Vol.2}, p.1654.
\bibitem{8}
P.~Di~Vecchia, Surveys in High Energ.Phys. 
{\bf 10}, 119 (1997); hep-th/9608090. 
\bibitem{8i}
P.~Di~Vecchia, {\it Duality in N=2,4 Supersymmetric Gauge Theories}, arXiv: hep-th/9803026.
\bibitem{9}
H.M.~Chan, J.~Faridani and S.T.~Tsou, Phys.Rev. {\bf D52}, 6134 (1995); 
arXiv: hep-th/9503106.
\bibitem{9i}
H.M.~Chan and S.T.~Tsou, Phys.Rev. {\bf D56}, 3646 (1997); arXiv: hep-th/9702117.
\bibitem{9ii}
H.M.~Chan and S.T.~Tsou, Phys.Rev. {\bf D57}, 2507 (1998); arXiv: hep-th/9701120.
\bibitem{10}
H.M.~Chan, J.~Faridani and S.T.~Tsou, Phys.Rev. {\bf D51}, 7040 (1995).
\bibitem{10i} 
H.M.~Chan, J.~Faridani and S.T.~Tsou, Phys.Rev. {\bf D53}, 7293 (1996); 
arXiv: hep-th/9512173.
\bibitem{11}
S.T.~Tsou, Int.J.Mod.Phys. {\bf A18S2}, 1 (2003); arXiv: hep-th/0110256.
\bibitem{12}
J.~Bordes, H.M.~Chan and S.T.~Tsou, {\it Updates of the dualized Standard Model in 
fermion masses and mixings}, arXiv: hep-ph/0302199.
\bibitem{13}
F.~Fucito, M.~Martellini and M.~Zeni, Nucl.Phys. {\bf B496}, 259 (1997); arXiv: hep-th/9605018.
\bibitem{14}
A.~Noguchi and A.~Sugamoto, VESTNIK of the Tomsk State Pedagogical University, Special isuue,  
{\bf 44N7}, 59 (2004); arXiv: hep-th/0408045.
\bibitem{15}
A.M.~Polyakov, Nucl.Phys. {\bf B164}, 171 (1980).
\bibitem{16}
D.~Zwanziger, Phys.Rev. {\bf 176}, 1489 (1968).
\bibitem{16i}
D.~Zwanziger, Phys.Rev. {\bf D3}, 880 (1971).
\bibitem{17}
R.A.~Brandt, F.~Neri and D.~Zwanziger, Phys.Rev. {\bf D19}, 1153 (1979).
\bibitem{18}
F.V.~Gubarev, M.I.~Polikarpov and V.I.~Zakharov, Phys.Lett. {\bf B438}, 147 (1998);
arXiv: hep-th/9805175.
\bibitem{19}
L.V.~Laperashvili and H.B.~Nielsen, Mod.Phys.Lett. {\bf A14}, 2797 (1999); 
arXiv: hep-th/9910101.
\bibitem{20}
G.~t' Hooft, Acta Physica Austrica suppl. {\bf XXII}, 531 (1980).
\bibitem{21}
C.~Montonen and D.~Olive, Phys.Lett. {\bf B72}, 117 (1977).
\bibitem{22}
M.~Gell--Mann and F.E.~Low, Phys.Rev. {\bf 95}, 1300 (1954).
\bibitem{23}
C.R.~Das and L.V.~Laperashvili, {\it Phase transition in the Higgs model of scalar dyons}, 
to be published in Mod.Phys.Lett. {\bf A}, (2006); arXiv: hep-ph/0511067.
\bibitem{24}
O.V.~Tarasov, A.A.~Vladimirov and A.Yu.~Zharkov, Phys.Lett. {\bf B93}, 429 (1980).
\bibitem{25}
Yu.A.~Simonov, Phys.Atom.Nucl. {\bf 58} 107 (1995) [Yad.Fiz. {\bf 58}, 113 (1995)];
arXiv: hep-ph/9311247.
\bibitem{25i}
A.M.~Badalian and Yu.A.~Simonov, Phys.Atom.Nucl. {\bf 60}, 630 (1997) 
[Yad.Fiz. {\bf 60}, 714 (1997)].
\bibitem{25ii}
A.M.~Badalian and D.S.~Kuzmenko, Phys.Rev. {\bf D65}, 016004 (2002); arXiv: hep-ph/0104097.
\bibitem{26}
I.M.~Narodetskii and M.A.~Trusov, Phys.Atom.Nucl. {\bf 65}, 917 (2002) 
[Yad.Fiz. {\bf 65}, 949 (2002)]; arXiv: hep-ph/0104019.
\bibitem{26i}
I.M.~Narodetskii and M.A.~Trusov, Phys.Atom.Nucl. {\bf 67}, 762 (2004)
[Yad.Fiz. {\bf 67}, 783 (2004)]; arXiv: hep-ph/0307131.
\bibitem{27}
A.A.~Abrikosov, Sov.Phys.JETP. {\bf 5}, 1174, (1957)
[Zh.Eksp.Teor.Fiz. {\bf 32}, 1442 (1957)].
\bibitem{28}
H.B.~Nielsen and P.~Olesen, Nucl.Phys. {\bf B61}, 45 (1973).
\bibitem{29}
L.V.~Laperashvili, D.A.~Ryzhikh and H.B.~Nielsen,
Int.J.Mod.Phys. {\bf A18}, 4403 (2003); arXiv: hep-th/0211224.
\bibitem{30}
C.R.~Das and L.V.~Laperashvili, Int.J.Mod.Phys. {\bf A20}, 5911 (2005); arXiv: hep-ph/0503138.
\bibitem{31}
G. 't~Hooft, Nucl.~Phys.~{\bf B190}, 455 (1981).
\bibitem{32}
A.S.~Kronfeld, G.~Schierholz and U.J.~Wiese, Nucl.Phys. {\bf B293}, 461 (1987).
\bibitem{33}
Yu.A.~Simonov, Phys.Usp. {\bf 39}, 313 (1996) [Usp.Fiz.Nauk. {\bf 166}, 337 (1996)];
arXiv: hep-ph/9709344.
\bibitem{34}
L.D.~Faddeev and A.J.~Niemi, Phys.Lett. {\bf B525}, 195 (2002); arXiv: hep-th/0101078.
\bibitem{35}
M.N.~Chernodub and F.V.~Gubarev, JETP Lett. {\bf 62}, 100 (1995); arXiv: hep-th/9506026.
\bibitem{36}
G.S.~Bali, C.~Schlighter and K.~Shilling, Prog.Theor.Phys.Suppl. {\bf 131}, 645 (1998); 
arXiv: hep-lat/9802005.
\bibitem{37}
M.N.~Chernodub, F.V.~Gubarev and M.I.~Polikarpov, JETP Lett. {\bf 69}, 169 (1999);
arXiv: hep-lat/9801010.
\bibitem{38}
M.N.~Chernodub, Phys.Rev. {\bf D69}, 094504 (2004); arXiv: hep-lat/0308031.
\bibitem{39}
T.~Suzuki, Prog.Theor.Phys. {\bf 80}, 929 (1988).
\bibitem{39i}
S.~Maedan and T.~Suzuki, Prog.Theor.Phys. {\bf 81}, 229 (1989).
\bibitem{40}
M.N.~Chernodub and M.I.~Polikarpov,  {\it Abelian projections and monopoles} in: 
``NATO Advanced Study Institute on Confinement, Duality and Non--perturbative Aspects of 
QCD'', Cambridge, England, 23 Jun - 4 Jul 1997, p.387: Ed. by Pierre van Baal, 
Plenum Press, 1998; arXiv: hep-th/9710205.
\bibitem{41}
M.N.~Chernodub, F.V.~Gubarev, M.I.~Polikarpov and A.I.~Veselov, 
Prog.Theor.Phys.Suppl. {\bf 131}, 309 (1998); arXiv: hep-lat/9802036.
\bibitem{42} 
M.N.~Chernodub, F.V.~Gubarev, M.I.~Polikarpov and V.I.~Zakharov, Phys.Atom.Nucl. {\bf 64}, 561 
(2001) [Yad.Fiz. {\bf 64}, 615 (2001)]; arXiv: hep-th/0007135.
\bibitem{43} 
M.N.~Chernodub, F.V.~Gubarev, M.I.~Polikarpov and V.I.~Zakharov, Nucl.Phys. 
{\bf B600}, 163 (2001);
arXiv: hep-th/0010265. 
\bibitem{43i} 
M.N.~Chernodub, F.V.~Gubarev, M.I.~Polikarpov and V.I.~Zakharov, Nucl.Phys. 
{\bf B592}, 107 (2001);
arXiv: hep-th/0003138.
\bibitem{44}
L.V.~Laperashvili and H.B.~Nielsen, {\it Multiple Point Principle and phase transition in gauge 
theories}, in: Proceedings of the International Workshop on ``What Comes Beyond the 
Standard Model'', Bled, Slovenia, 29 June - 9 July, 1998 (DMFA, Zaloznistvo, Ljubljana, 1999), 
p.15; arXiv: hep-ph/9905357.
\bibitem{44i}
L.V.~Laperashvili and H.B.~Nielsen, 
{\it Phase transition couplings in the Higgsed monopole model},
arXiv: hep-th/9909181.
\bibitem{45}
L.V.~Laperashvili and H.B.~Nielsen, Int.J.Mod.Phys. 
{\bf A16}, 2365 (2001); arXiv: hep-th/0010260.
\bibitem{46}
L.V.~Laperashvili, H.B.~Nielsen and D.A.~Ryzhikh, Int.J.Mod.Phys. 
{\bf A16}, 3989 (2001); arXiv: 
hep-th/0105275.
\bibitem{47}
L.V.~Laperashvili, H.B.~Nielsen and D.A.~Ryzhikh, Phys.Atom.Nucl. 
{\bf 65}, 353 (2002) [Yad.Fiz. {\bf 65}, 377 (2002)]; arXiv: hep-th/0109023.
\bibitem{48}
L.V.~Laperashvili and D.A.~Ryzhikh, {\it Multiple point model and phase 
transition couplings in the two--loop approximation of dual scalar electrodynamics} in:
Proceedings of the International Workshop on ``What Comes Beyond the Standard Model'', 
{\bf Vol.2}, Bled, Slovenia, 17 - 27 July, 2001 (DMFA, Zaloznistvo, Ljubljana, 2002), p.131; 
arXiv: hep-ph/0112183.
\bibitem{49}
L.V.~Laperashvili and D.A.~Ryzhikh, 
{\it Phase transition in gauge theories and the Planck scale 
physics}, preprint {\bf ITEP--24--01}, October, 2001, 82pp; arXiv: hep-ph/0212221.
\bibitem{50}
M.N.~Chernodub, R.~Feldmann, E.M.~Ilgenfritz and A.~Schiller, 
Phys.Rev. {\bf D71}, 074502 (2005); arXiv: hep-lat/0502009
\bibitem{51}
G.~Schierholz, {\it On the structure of the Yang-Mills vacuum}, in: Proceedings of the 
International RCNP Workshop on ``Colour Confinement and Hadrons'', Osaka, Japan, March, 1995; 
Ed. by H.~Toki et. al. (World Scientific, Singapore, 1995), p.96; arXiv: hep-lat/9506033.
\bibitem{51a}
V.~Bornyakov and G.~Schierholz, Phys.Lett. {\bf B384}, 190 (1996); arXiv: hep-lat/9605019.
\bibitem{51ai} 
V.~Bornyakov and G.~Schierholz, Nucl.Phys.Proc.Suppl. {\bf 53}, 484, (1997).
\bibitem{51b}
E.T.~ Akhmedov, M.N.~ Chernodub and M.I.~ Polikarpov, JETP Lett. {\bf 67}, 389 (1998); arXiv: 
hep-th/9802084.
\bibitem{52}
M.N.~Chernodub, F.V.~Gubarev and M.I.~Polikarpov, JETP Lett. {\bf 69}, 169 (1999); arXiv: 
hep-lat/9801010.
\bibitem{53}
N.N.~Bogolyubov and D.V.~Shirkov, Doklady AN SSSR (Reports of AS USSR) {\bf 103}, 203 (1955).
\bibitem{53i}
N.N.~Bogolyubov and D.V.~Shirkov, Doklady AN SSSR (Reports of AS USSR) {\bf 103}, 391 (1955). 
\bibitem{53ii}
N.N.~Bogolyubov and D.V.~Shirkov, JETP {\bf 30}, 77 (1956).
\bibitem{54}
L.D.~Landau, A.A.~Abrikosov and I.M.~Khalatnikov, Doklady AN SSSR
(Reports of AS USSR) {\bf 95}, 773 (1954).
\bibitem{54i}
L.D.~Landau, A.A.~Abrikosov and I.M.~Khalatnikov, Doklady AN SSSR
(Reports of AS USSR) {\bf 95}, 1177 (1954).
\bibitem{55}
J.~Jersak, T.~Neuhaus and P.M.~Zerwas, Phys.Lett. {\bf B133}, 103 (1983). 
\bibitem{55i}
J.~Jersak, T.~Neuhaus and P.M.~Zerwas, Nucl.Phys. {\bf B251}, 299 (1985).
\bibitem{56}
J.~Jersak, T.~Neuhaus and H.~Pfeiffer, Phys.Rev. {\bf D60}, 054502 (1999);
 arXiv: hep-lat/9903034.
\bibitem{56a}
L.V.~Laperashvili and H.B.~Nielsen, Mod.Phys.Lett. {\bf A12}, 73 (1997).
\bibitem{57}
L.V.~Laperashvili and H.B.~Nielsen, {\it The problem of monopoles in the Standard and Family 
replicated models}, a talk given at the 11th Lomonosov Conference on ``Elementary Particle 
Physics'', Moscow, Russia, 21 - 27 August, 2003, published in: {\it Moscow 2003, 
Particle physics in laboratory, space and universe}, p.331; arXiv: hep-th/0311261.
\bibitem{58}
C.D.~Froggatt, L.V.~Laperashvili, H.B.~Nielsen and Y.~Takanishi, 
{\it Family replicated gauge group
models}, in: Proceedings of the Fifth International Conference ``Symmetry in Nonlinear 
Mathematical Physics'', Kiev, Ukraine, 23 - 29 June, 2003, Ed. by A.G.~Nikitin, V.M.~Boyko, 
R.O.~Popovich and I.A.~Yehorchenko (Institute of Mathematics of NAS of Ukraine, Kiev, 2004), 
{\bf Vol.50}, Part 2, p.737; arXiv: hep-ph/0309129.
\bibitem{59}
L.V.~Laperashvili, Phys.Atom.Nucl. {\bf 57}, 471 (1994) [Yad.Fiz. {\bf 57}, 501 (1994)].
\bibitem{59i}
L.V.~Laperashvili, Phys.Atom.Nucl. {\bf 59}, 162 (1996) [Yad.Fiz. {\bf 59}, 172 (1996)].
\bibitem{60}
P.A.M.~Dirac, Proc.Roy.Soc.Lond. {\bf A133}, 60 (1931).
\bibitem{61}
P.A.~Kovalenko and L.V.~Laperashvili, Phys.Atom.Nucl. {\bf 62}, 1729 (1999)
[Yad.Fiz. {\bf 62}, 1857 (1999)].
\bibitem{62}
P.A.~Kovalenko and L.V.~Laperashvili, 
{\it The effective QCD Lagrangian and renormalization group 
approach}, {\it ITEP--PH--11--97}, August, 1997, 5pp. Talk given at ``8th Lomonosov Conference 
on Elementary Particle Physics'', Moscow, Russia, 25 - 29 August, 1997; arXiv: hep-ph/9711390.
\bibitem{63}
S.G.~Matinyan and G.K.~Savvidy, Nucl.Phys. {\bf B134}, 539 (1978).
\bibitem{64}
S.G.~Matinyan and G.K.~Savvidy, Yad.Fiz. {\bf 25}, 218 (1977).
\bibitem{65}
I.A.~Batalin, S.G.~Matinyan and G.K.~Savvidy, Sov.J.Nucl.Phys. {\bf 26}, 214 (1977) 
[Yad.Fiz. {\bf 26}, 407 (1977)].
\bibitem{66}
V.V.~Vladimirsky, Phys.Atom.Nucl. {\bf 65}, 305 (2002) [Yad.Fiz. {\bf 65}, 330 (2002)]. 
\bibitem{66i}
V.V.~Vladimirsky, Phys.Atom.Nucl. {\bf 66}, 2214 (2003) [Yad.Fiz. {\bf 66}, 2266 (2003)].
\bibitem{67}
M.A.~Shifman, A.I.~Vainstein and V.I.~Zakharov, Nucl.Phys. {\bf B147}, 385 (1979).  
\bibitem{67i}
M.A.~Shifman, A.I.~Vainstein and V.I.~Zakharov, Nucl.Phys. {\bf B147}, 448 (1979).
\bibitem{68}
V.A.~Novikov, M.A.~Shifman, A.I.~Vainstein, M.B.~Voloshin and V.I.~Zakharov, Nucl.Phys. 
{\bf B237}, 525 (1984).
\bibitem{69}
H.B.~Nielsen and P.~Olesen, Nucl.Phys. {\bf B160}, 380 (1979).
\bibitem{70}
C.R.~Das, L.V.~Laperashvili and H.B.~Nielsen, {\it Generalized dual symmetry of nonabelian 
theories, monopoles and dyons}, a talk given at the 12th Lomonosov Conference on 
``Elementary Particle Physics'', Moscow, 25 - 31 August, 2005; arXiv: hep-ph/0510392.
\bibitem{71}
S.~Coleman and E.~Weinberg, Phys.Rev. {\bf D7}, 1888 (1973).
\bibitem{72}
M.~Sher, Phys.Rept. {\bf 179}, 273 (1989).
\bibitem{73}
D.~Gross, in: ``Methods of Field Theory'', Proc. 1975 Les Houches Summer 
School, eds R.~Balian and J.~Zinn-Justin, North Holland, Amsterdam, 1975.
\end{thebibliography}
\end{document}